\documentclass[3p,times]{elsarticle}
\usepackage{hyperref,multirow,graphicx,booktabs}
\usepackage[usenames,dvipsnames]{color}
\journal{arXiv}

\graphicspath{{fig/}}
\biboptions{authoryear}
\begin{document}

\begin{frontmatter}

\title{Collective Learning in China's Regional Economic Development}
\author[inst1,inst2,inst3]{Jian Gao}
\author[inst2]{Bogang Jun}
\author[inst2]{Alex ``Sandy'' Pentland}
\author[inst1,inst3]{Tao Zhou}
\author[inst2]{C{\'e}sar A. Hidalgo\corref{cor1}}
\cortext[cor1]{\emph{Email address}: hidalgo@mit.edu}

\address[inst1]{CompleX Lab, University of Electronic Science and Technology of China, Chengdu 611731, China}
\address[inst2]{MIT Media Lab, Massachusetts Institute of Technology, Cambridge, MA 02139, USA}
\address[inst3]{Big Data Research Center, University of Electronic Science and Technology of China, Chengdu 611731, China}

\begin{abstract}
Industrial development is the process by which economies learn how to produce new products and services. But how do economies learn? And who do they learn from? The literature on economic geography and economic development has emphasized two learning channels: inter-industry learning, which involves learning from related industries; and inter-regional learning, which involves learning from neighboring regions. Here we use 25 years of data describing the evolution of China's economy between 1990 and 2015--a period when China multiplied its GDP per capita by a factor of ten--to explore how Chinese provinces diversified their economies. First, we show that the probability that a province will develop a new industry increases with the number of related industries that are already present in that province, a fact that is suggestive of inter-industry learning. Also, we show that the probability that a province will develop an industry increases with the number of neighboring provinces that are developed in that industry, a fact suggestive of inter-regional learning. Moreover, we find that the combination of these two channels exhibit diminishing returns, meaning that the contribution of either of these learning channels is redundant when the other one is present. Finally, we address endogeneity concerns by using the introduction of high-speed rail as an instrument to isolate the effects of inter-regional learning. Our differences-in-differences (DID) analysis reveals that the introduction of high speed-rail increased the industrial similarity of pairs of provinces connected by high-speed rail. Also, industries in provinces that were connected by rail increased their productivity when they were connected by rail to other provinces where that industry was already present. These findings suggest that inter-regional and inter-industry learning played a role in China's great economic expansion.
\end{abstract}


\end{frontmatter}

%---------------------------------------------------------------------------
% \section*{Introduction}
%---------------------------------------------------------------------------
\section{Introduction}

Between 1990 and 2015 China experienced one of the fastest episodes of economic growth in our recorded history. China's overall GDP grew by a factor of 30, from less than USD 400 billion in 1990 to more than USD 10 trillion in 2015. The per capita economic growth of China was also outstanding. China's GDP per capita, adjusted by purchasing power parity (PPP) and at constant prices, increased by nearly a factor of 10, from USD 1,516 in 1990 to more than USD 13,400 in 2015. For comparison, in the same period global GDP grew only by a factor of three (from USD 22.5 trillion to USD 73.4 trillion) and global GDP per capita, also at PPP and constant prices, grew by less than a factor of two (from USD 8,876 to USD 14,602).

The pace and scale of China's great economic expansion have no historical precedent \citep{Song2011,Zhu2012,Eichengreen2012,Felipe2013}. If China's GDP per capita, at PPP and constant prices, continued growing at the same pace, it would surpass USD 130,000 per capita by the year 2040. But China is unlikely to repeat this success in the next 25 years. This suggests that China's great expansion was probably a distinct event in economic history, and one from which many countries could learn.

But what explains China's remarkable economic success? One theory is that China's great expansion relied on the export of products that were unusually sophisticated for China's level of income \citep{Rodrik2006, Hidalgo2009, Hausmann2014, Hidalgo2015}. During this period China exported products like electronics and other advanced manufactures that were at that time being produced mostly in countries with an income per capita that was much larger than that of China \citep{Lin2012}. By succeeding in the export of these sophisticated products, China was able to penetrate markets that could support higher wages, and consequently, higher incomes.

Evidence in support of this theory is shown in the work of \cite{Rodrik2006}, who estimated the level of sophistication of Chinese exports by calculating the average income per capita of the countries exporting the same products than China. \cite{Rodrik2006} showed that even as early as 1992, when China's GDP per capita at PPP and constant prices was just USD 1,844, it exported products associated with an average level of income that was roughly of USD 13,500, which corresponds to China's level of income in 2015. \cite{Rodrik2006} argued that this unusually high level of export sophistication fueled China's great economic expansion.

Further evidence supporting the idea that the sophistication of China's exports is a factor explaining China's rapid economic expansion is contained in the literature on economic complexity \citep{Hidalgo2009, Tacchella2012, Hausmann2014, Hidalgo2015}, which has focused on developing measures of a country's export sophistication that avoid the circularity of using income data. The consensus of this literature is also that countries with a relatively high level of economic complexity--countries that export a diverse set of non-ubiquitous goods--grow, on average, faster than countries with a similar level of income but lower levels of economic complexity.

If China's economy expanded because it succeeded in the export of sophisticated products, then the question is: how did China learn to produce products of increasing levels of sophistication? Here the literature provides two answers. One is that economies learn by leveraging the capabilities embodied in related industries. That is economies that are good at producing shirts, would have an easier time learning how to produce pants, coats, and socks. The other idea is that economies learn from neighboring regions. That is the probability that a province would succeed at making shirts depends on having neighboring regions that have already developed the capacity to produce shirts.

The view that economic development is a collective learning process is found repeatedly in the work of development economists, evolutionary economists, economic geographers, and in the literature of economic clusters.

Evolutionary economists, going back to the seminal work of \cite{Nelson1982}, have pushed the idea that economies learn by accumulating capabilities in networks of individuals and firms. In this strand of literature, capabilities are explicit and tacit knowledge \citep{Polanyi1958,Collins2010} that firms embody in routines and procedures that make the learning process deeply path dependent. The ability of a firm to accumulate these capabilities depends, among other factors, on the institutional environment of where the firm is located \citep{Saxenian1996}, the levels of trust in the population \citep{Fukuyama1995}, the firms' organizational structure \citep{Powell1990}, the dynamic capacity of a firm to learn \citep{Teece1994b}, the social networks where the economy is embedded \citep{Granovetter1985}, and of course, on the existence of related firms and neighboring regions that have already accumulated the right capabilities.

It is not surprising, therefore, that much work has gone into understanding the channels that facilitate the ability of economies to learn. In broad strokes, this literature has focused on two learning channels: inter-industry learning, which has been studied extensively, and focuses on how firms learn from related industries that are already in their region; and inter-regional learning, which has been much less studied, and focuses on the learning that takes place across geographic boundaries.

Inter-industry learning has been studied at the international, regional, and firm level. This literature has focused on testing how the existence of related industries increases the probability that an industry will enter a region, exit a region, or became more productive.

At the international level, \cite{Hidalgo2007} and \cite{Hausmann2014} have used export data to show that the probability that a country will develop comparative advantage in a new product depends strongly on the number of related products that it already exports. To establish this stylized fact \cite{Hidalgo2007} introduced the idea of the \emph{product space}, a network connecting products that countries are likely export in tandem. Using this network representation it is easy to score each product that a country does not yet export based on the number of related products that the same country is already exporting. This score, called \emph{density}, is a statistically good predictor of the probability that a country will develop comparative advantage in a specific product in the future.

At the regional level, people have used data on input-output relationships, labor flows, and the product portfolios of manufacturing plants to measure industrial relatedness \citep{Boschma2017, Delgado2016, Boschma2012, Semitiel2012, Boschma2009, Frenken2007}. \cite{Neffke2011} used data on the product portfolios of manufacturing plants in Sweden to connect industries and showed respectively that the probability that an industry will enter, or exit, a region, increases, or decreases, with the number of related industries present in it. \cite{Delgado2014} used data from the US Cluster Mapping Project to show that firms located in clusters of related industries tend to experience higher patenting and employment growth. \cite{Delgado2010} also shows that clusters tend to enhance entrepreneurship, since start-up industries located in clusters tend to grow faster.

At the firm level, \cite{Teece1980} has argued that coherent multi-product enterprises (firms that produce a diverse portfolio of related products) are an efficient way to organize economic activity when the development of products requires re-utilizing proprietary knowhow and specialized and indivisible physical assets \citep{Teece1982}. More recently, empirical work like that of \cite{Neffke2013} has leveraged labor-flow data to connect related industries, by arguing that industries that are more likely to exchange labor are related in terms of the skills that they require. Using their skill-relatedness metric, they found that firms are more likely to diversify their product portfolios to include the products that were being produce by related industries \citep{Neffke2013}, adding to the evidence that firm diversification is also path dependent and coherent \citep{Teece1994}. By studying empirically the effects of five different dimensions of agglomeration on the survival chances of new entrepreneurial firms in China, \cite{Howell2016} found that increasing local related variety has a stronger positive effect on new firm survival than other types of agglomeration.

The inter-regional learning literature, on the other hand, is sparser, and it focuses on how economies learn from neighboring regions instead of similar industries. At the international level, \cite{Bahar2014} showed that the probability that a country will start exporting a product increases significantly if that country shares a border with a neighbor that is already a successful exporter of that product, even after discounting the effects of product relatedness captured in the product space. At the regional level, \cite{Boschma2016} used data from the United States to show that regions are more likely to develop industries that are present in neighboring regions. \cite{Acemoglu2015} studied the direct and spillover effects of local state capacity in Colombia, and found that spillover effects are sizable, accounting for about 50 percent of the quantitative impact of an expansion in local state capacity. At the firm level, \cite{Holmes2011} studied the geographic expansion of Wal-Mart stores in the US, and found that locations of new Wal-Mart stores tend to be in close geographic proximity to regions where Wal-Mart already had a high density of stores.

Yet, one of the issues that limits this literature is the sparse causal evidence supporting both inter-industry learning and inter-regional learning. One effort in this direction is the work of \cite{Ellison2010}, who explored data from US manufacturing industries to check the effect of the cost of moving goods, people, and ideas on the co-location of industries, i.e. inter-industry learning. To reduce concerns of reverse causality, \cite{Ellison2010} used data from UK industries and from US areas as instruments.

In this paper, we contribute to this expanding body of literature by studying the role of collective learning in the great economic expansion experienced by China between 1990 and 2015. First, we show that the probability that a new industry will grow in a province increases with the number of related industries present in it, a fact supporting theories of inter-industry learning. Next, we show that the probability that a new industry will grow in a province also increases with the number of neighboring provinces in which that industry is already present, a fact that supports inter-regional learning theories. Moreover, we find that both learning channels work together but exhibit diminishing returns, meaning that when one learning channel is sufficiently active (inter-industry or inter-regional) the marginal contribution of the other one is reduced (the channels are substitutes). Finally, we address endogeneity concerns by using the introduction of high-speed rail among Chinese provinces to isolate the effects of inter-regional learning. The introduction of high-speed rail is an instrument that affects the travel time between provinces, but not the similarity among industries. Our differences-in-differences (DID) results show that, after the introduction of high speed rail, the pairs of provinces connected by rail became more similar in terms of their industrial structure. Also, our results show a significant increase in productivity for industries located in provinces that became connected by high-speed rail to other provinces where that industry was present. Together, these results add to the evidence that China's economic expansion benefited from inter-industry and inter-regional learning.

%---------------------------------------------------------------------------
% \section*{Data}
%---------------------------------------------------------------------------
\section{Data}

\subsection{China's firm data}

We use data from China's stock market extracted from the RESSET Financial Research Database, which is provided by Beijing Gildata RESSET Data Tech Co., Ltd.\footnote{http://www.resset.cn}, a leading provider of economic and financial data in China. Our data set covers 1990-2015, a period when China achieved rapid economic development. This data set provides some basic registration and financial information of publicly listed firms in Chinese stock exchanges, such as listing date, delisting date, registered address, industry category, yearly revenue, and number of employees. Although the numbers of newly listed and delisted firms in each year fluctuates, the overall number of firms increases almost linearly with time (see Figure~\ref{FigS1:Stat}). The registered addresses of firms cover 31 provinces in China. All these listed firms in our data set are aggregated into two levels, 18 categories at the sector level and 70 subcategories at the sub-sector level. The aggregation is based on the ``Guidelines for the Industry Classification of Listed Companies'' issued by the China Securities Regulatory Commission (CSRC)\footnote{http://www.csrc.gov.cn} in 2011 (see online Appendix for details). CSRC category and CSRC subcategory codes as well as their associated industry names are shown in Figure~\ref{FigS2:CSRC}. Moreover, to measure the productivity of the industries in a province, we use the total revenue of firms divided by the total number of employees in that industry in that province.

\subsection{Distance and macroeconomic indicators}
To study inter-regional learning, we use geographic distance as a measure of physical proximity between regions. The geographic distance ($D_{i,j}$) between provinces $i$ and $j$ is defined as the distance between the capital cities of two provinces (in China capital cities are likely to be the largest city in a province). Together, we collect macroeconomic data at the province-level, including Gross Domestic Product per capita (GDP per capita), resident population, total value of imports and exports, urban area, and total area, from the China Statistical Yearbook, which is published by the National Bureau of Statistics of China\footnote{http://www.stats.gov.cn}. As an urbanization metric, we use the share of urban area in a province. These macroeconomic indicators cover the 1990-2015 period and are available for the 31 provinces in China. Brief descriptions and summary statistics of these distance and macroeconomic indicators can be found in Table~\ref{Tab:Unit}.

%---------------------------------------------------------------------------
% \section*{Results}
%---------------------------------------------------------------------------
\section{Results}

We organize our results into four sections. First, we explore inter-industry learning by constructing a network of related industries, or industry space, and explore how the probability that an industry will emerge in a province increases with the number of related industries already present in it. Next, we explore inter-regional learning by using geographic data to study how the probability that an industry will emerge in a province increases with the presence of that industry in neighboring provinces. Then, we combine both geographic and industrial similarity data to study the interaction between inter-industry and inter-regional learning. Our results shows that learning exhibits diminishing returns, meaning that when one learning channel (inter-regional or inter-industry) is sufficiently active, the other channel does not contribute as much (they are substitutes). Finally, we use the introduction of high-speed rail between provinces as an instrument to gather evidence in support of the inter-regional learning hypothesis. Our differences-in-differences (DID) estimate shows that the provinces connected by high speed rail experienced a significant increase in their industrial similarity, and also, that industries located in provinces that were connected by rail increased their productivity when these rail connections connected them to other provinces where the same industry was present.

\subsection{Inter-industry learning}

We explore how the probability that an industry will appear in a province is affected by the number of related industries already present in it, by first constructing a network of industries, or industry space, and then use this network to see if the probability that an industry will emerge in a province increases with the number of related industries that are already present in it.

First, we connect provinces and industries by building a ``province-industry'' bipartite network, where the weight of link $x_{i,\alpha}$ is the number of firms in province $i$ that operate in industry $\alpha$ (see Figure~\ref{FigS3:BiNet} for illustration). Going forward, we use Greek letters for indices indicating industries and Roman letters for indices indicating provinces.

Next, we estimate the proximity $\phi_{\alpha,\beta}$ between industries $\alpha$ and $\beta$ by calculating the cosine similarity between $x_{i,\alpha}$ and $x_{i,\beta}$ across all provinces. Following \cite{Hidalgo2007}, we assume the co-location of industries to be an imperfect proxy of their similarity, since pairs of industries that tend to co-locate are more likely to require similar capabilities (whether these are human capital, institutional factors, logistic facilities, or geographic resources) than pairs of industries that do not tend to be co-located. Formally, we let $x_{i,\alpha,t}$ and $x_{i,\beta,t}$ be the number of firms in province $i$ that respectively operate in industries $\alpha$ and $\beta$ at year $t$. Then, the proximity $\phi_{\alpha,\beta,t}$ is given by:
\begin{equation}
\phi_{\alpha,\beta,t} =\frac{ \sum_{i}{ x_{i,\alpha,t} x_{i,\beta,t} }}{\sqrt{\sum_{i}{(x_{i,\alpha,t})^2}} \sqrt{\sum_{i}{(x_{i,\beta,t})^2}}}.
\label{Eq:SimIndu}
\end{equation}

\linespread{1.5}
\begin{figure}[t]
  \centering
  \includegraphics[width=0.98\textwidth]{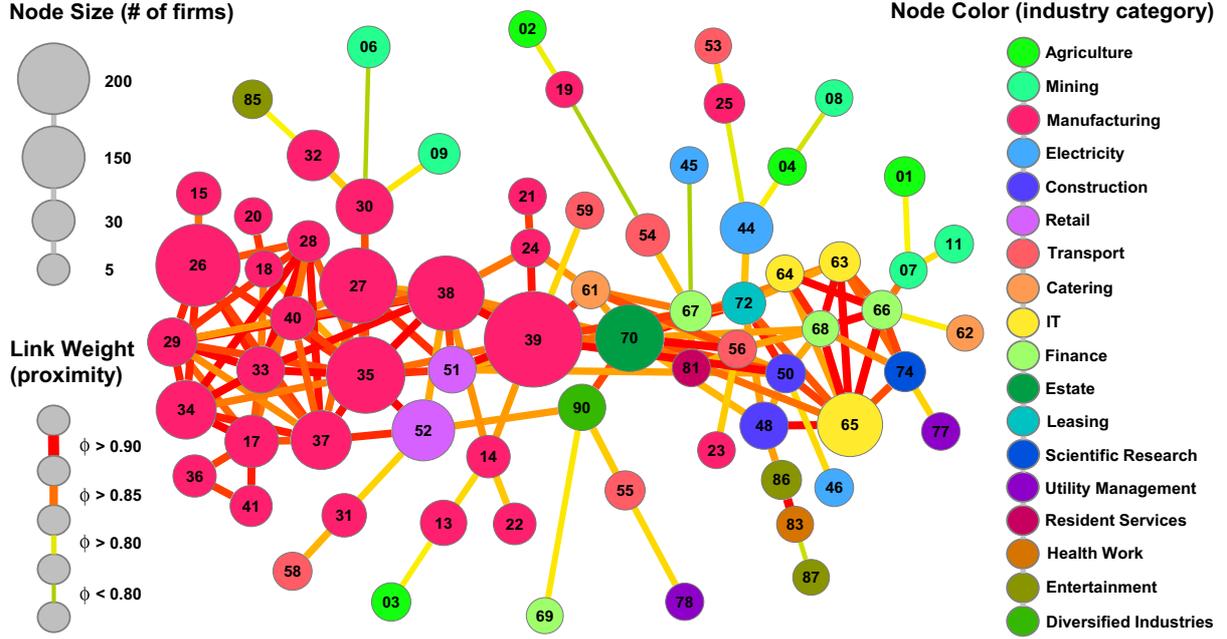}
  \caption{Network representation of China's industry space in 2015. Nodes (circles) represent industries. Links connect industries that are likely to locate in the same province. Nodes are classified into 70 subcategories and colored according to 18 sectors. The size of each node is proportional to the number of firms in that industry. The color and weight of links correspond to the proximity value ($\phi$) between two industries.}
  \label{Fig1:Space}
\end{figure}

Figure~\ref{Fig1:Space} shows China's industry space for the year 2015 (see Figure~\ref{FigS4:Net} and online Appendix for details on the visualization methods used). We note that China's industry space exhibits both, a core-periphery and a dumbbell structure, with a tightly knit core of manufacturing industries (on the left), and another tightly knit core of service and information related activities (on the right). This dumbbell structure is also visible when looking at the hierarchically clustered matrix of industrial proximities (see Figure \ref{FigS5:Modu}). In agreement with previous findings, which used data on products instead of industries \citep{Hidalgo2007}, we find extractive and agricultural activities to occupy the periphery of the industry space.

Next, we look at how the structure of the industry space shapes the economic diversification paths of Chinese provinces using three methods: a network visualization, a graphical method, and a multivariate statistical model.

First, we define an industry to be present in a province if that province has revealed comparative advantage in that industry. We define the revealed comparative advantage $RCA_{i,\alpha,t}$ for province $i$ in industry $\alpha$ at year $t$ following \cite{Balassa1965}. That is, we use the ratio between the observed number of firms operating in industry $\alpha$ in province $i$ and the expected number of firms of that industry in that province. Formally, the revealed comparative advantage $RCA_{i,\alpha,t}$ is given by:
\begin{equation}
    RCA_{i,\alpha,t} = \left.{\frac{x_{i,\alpha,t}}{\sum_{\alpha}x_{i,\alpha,t}}}\middle/ \frac{\sum_{i}x_{i,\alpha,t}}{\sum_{\alpha}\sum_{i}x_{i,\alpha,t}}\right. ,
    \label{Eq:RCA}
\end{equation}
where $x_{i,\alpha,t}$ is the number of firms in province $i$ that operate in industry $\alpha$ at year $t$. We say industry $\alpha$ is present in province $i$ at year $t$ if $RCA_{i,\alpha,t} \geq 1$.

\begin{figure}[!t]
  \centering
  \includegraphics[width=0.99\textwidth]{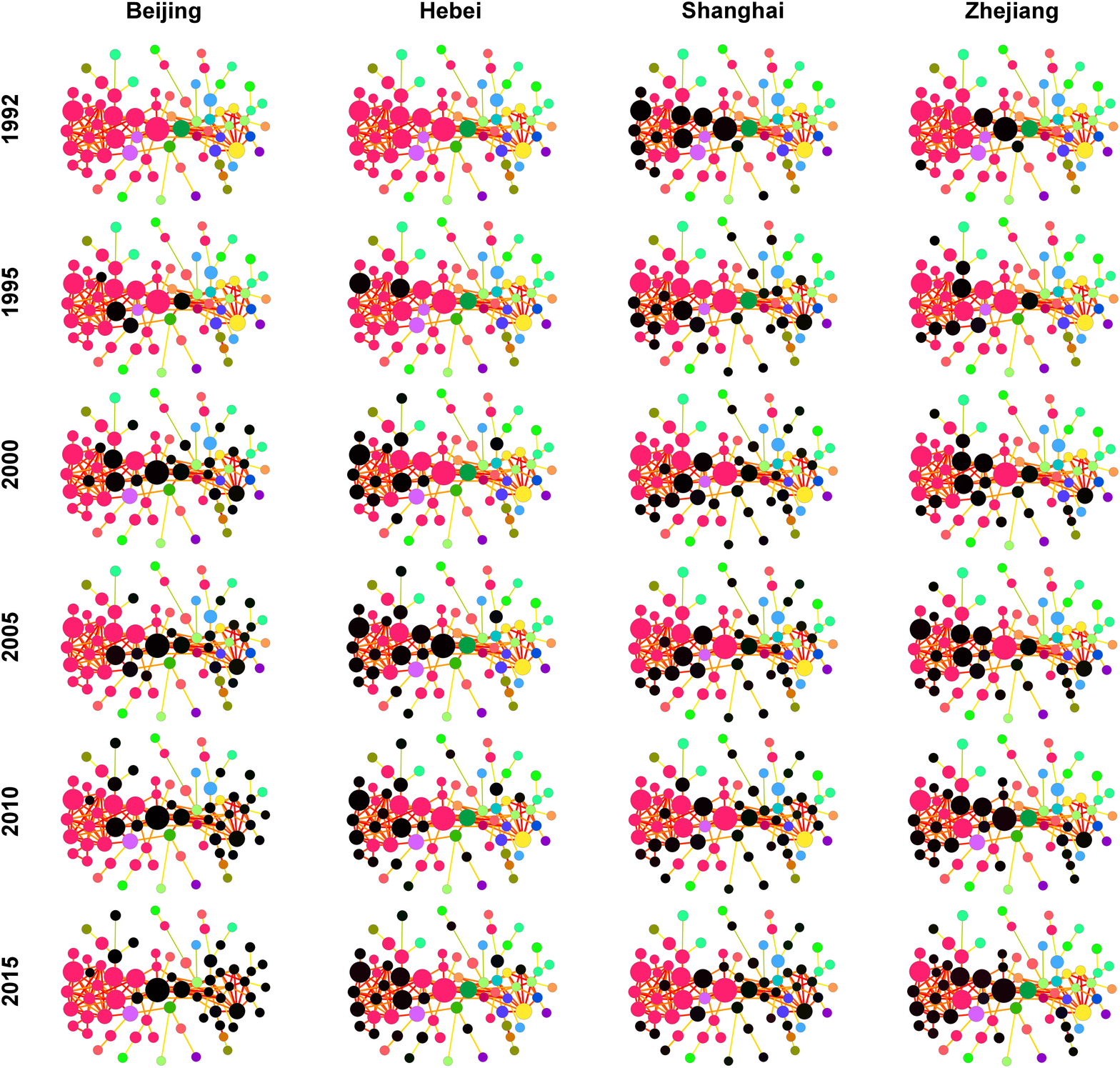}
  \caption{Evolution of China's provincial industrial structure between 1992 and 2015. Four illustrated provinces are Beijing, Hebei, Shanghai, and Zhejiang. Black circles indicate industries in which a province has revealed comparative advantage ($RCA\geq 1$).}
  \label{Fig2:InduEvo}
\end{figure}

Figure~\ref{Fig2:InduEvo} uses black circles to show the industries that were present in Beijing, Hebei, Shanghai, and Zhejiang, in 1992, 1995, 2000, 2005, 2010, and 2015. In these four illustrative examples, we can see that the new industries that are present in each of these provinces tend to be connected to other industries that were already present in that province. For example, Beijing and Shanghai gradually occupy Internet and financial services while Hebei and Zhejiang gradually occupy manufacturing industries. In the case of Shanghai, we see how the province gradually shifts its revealed comparative advantage from manufacturing to service and information activities during this period of economic development.

Next, we formalize this observation by constructing an indicator for each industry and province, counting the number of related industries that are already present in that province (i.e., $RCA\geq 1$). In the literature this estimator is called \emph{density} \citep{Hidalgo2007,Boschma2013,Boschma2016}. Here, to avoid confusion with a similar indicator we will introduce later for neighboring provinces, we call this estimator the density of active related industries ($\omega$). Formally, the density of active related industries ($\omega_{i,\alpha,t}$) for industry $\alpha$ in province $i$ at year $t$ is given by:
\begin{equation}
    \omega_{i,\alpha,t} =\frac{\sum_{\beta}{\phi_{\alpha,\beta,t} U_{i,\beta,t}}}{\sum_{\beta}{\phi_{\alpha,\beta,t}}},
    \label{Eq:Induden}
\end{equation}
where $U_{i,\beta,t}$ takes the value of 1 if province $i$ has revealed comparative advantage in industry $\alpha$ at year $t$ (i.e., $RCA_{i,\beta,t}\geq 1$) and 0 otherwise. Density is simply an indicator telling us, for each industry, what is the fraction of related industries that are already present in that province.

Next, we look at the probability that industry $\alpha$ would appear in province $i$ as a function of the density of active related industries in that province. To reduce noise, we follow \cite{Bahar2014} and restrict the appearance of new industries to two conditions: a backward condition, which requires an industry to have RCA below 1 during the two years prior to the beginning of the period; and a forward condition, which requires an industry to sustain RCA above 1 for the two years after the end of the period.

\begin{figure}[t]
  \centering
  \includegraphics[width=0.9\textwidth]{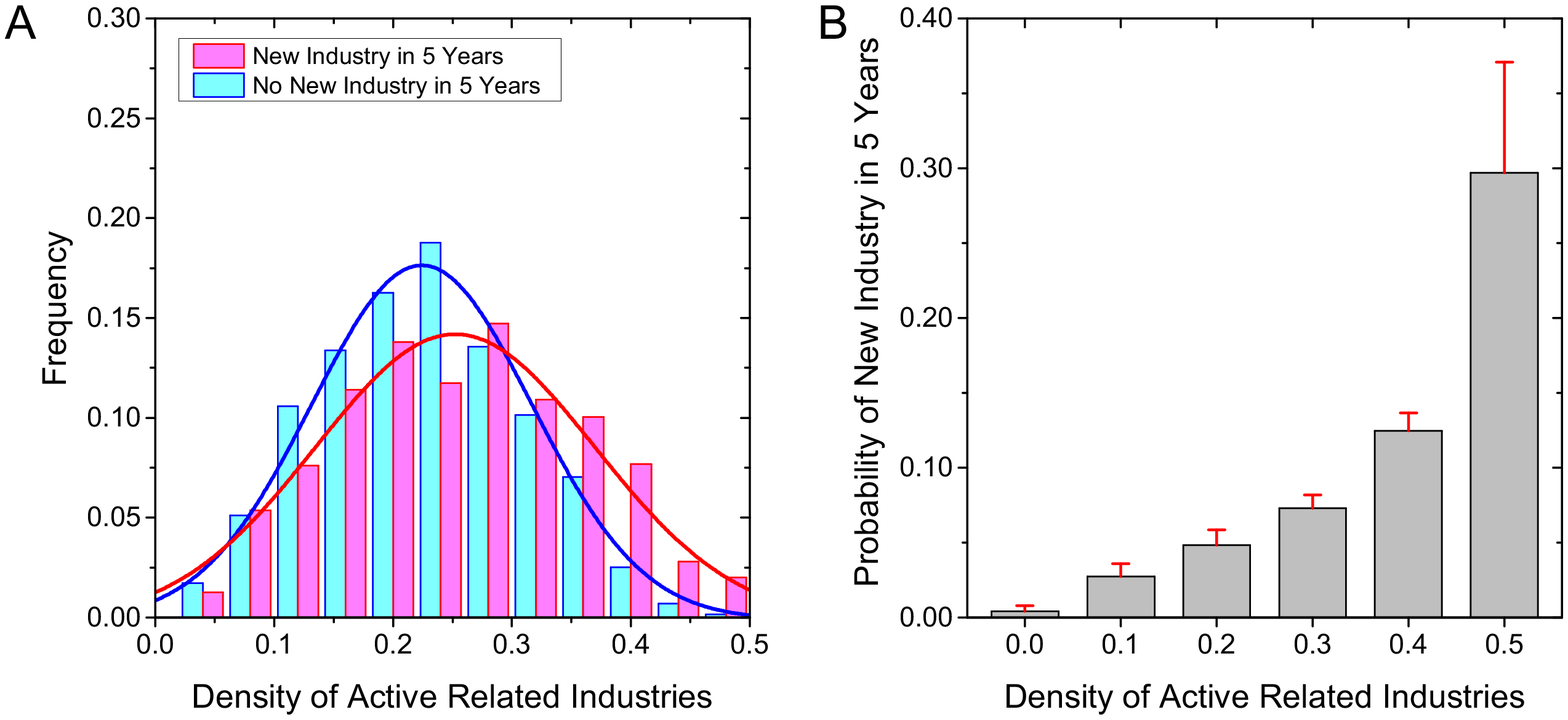}
  \caption{Inter-industry learning. (A) Distribution of the density of active related industries for each pair of provinces and industries. The pink distribution focuses only on pairs of provinces and industries that developed revealed comparative advantage in the next five years. The blue distribution is for the pairs of industries and provinces that did not develop revealed comparative advantage. The mean of the pink distribution is significantly larger than that of the blue distribution (ANOVA p-value=$2.1\times10^{-40}$). (B) Probability that a new industry will appear in a province as a function of the density of active related industries ($\omega$). Bars indicate average values and error bars indicate standard errors. Results show averages for 2001-2015 using five-year intervals. In all calculations, densities were calculated for the base year.}
  \label{Fig3:InduDen}
\end{figure}

Figure~\ref{Fig3:InduDen}A shows the frequency of densities of active related industry for pairs of industries and provinces that developed revealed comparative advantage (in pink) and that did not develop revealed comparative advantage (in blue) in a five-year period. The distributions show that--on average--the density of an industry in the provinces that developed revealed comparative advantage in that industry five years later is significantly larger (ANOVA p-value=$2.1\times10^{-40}$) than in those that did not (see Figure~\ref{FigS6:InduNum} for additional robustness check).

Figure~\ref{Fig3:InduDen}B looks at the probability that a province will develop revealed comparative advantage in an industry as a function of the density of active related industries in that province five years ago. The increasing and convex relationship shows that the probability that an industry will develop revealed comparative advantage in a province increases strongly with the density of active related industries. To reduce noise, we use a fixed industry space ($\phi_{\alpha,\beta}$) in year 2015 in Eq.~(\ref{Eq:Induden}), but we note that our results are robust (see Figure~\ref{FigS7:InduDen}) when we use a time-varying industry space ($\phi_{\alpha,\beta,t}$), where the industrial proximity is calculated using data only from previous years.

Finally, we use a multivariate probit model to estimate how the probability that a province will develop revealed comparative advantage, or keep revealed comparative advantage in an industry, changes with the density of active related industries. We separate our dataset into two sets: one set containing all province-industry pairs that did not have revealed comparative advantage (that could potentially be developed), and another set containing all pairs of provinces and industries that had comparative advantage (and that could lose it). Then, we set up two probit regressions, one explaining the probability that a province without RCA in an industry will develop RCA in that industry in the next five years, and the other explaining the probability that a province with RCA in an industry will keep RCA in that industry. In both of these regressions we control for the number of provinces with revealed comparative advantage in that industry and the number of industries with revealed comparative advantage in that province. Our empirical specification is:
\begin{equation}
    U_{i,\alpha,t+5} = \beta_{0} + \beta_{1}\omega_{i,\alpha,t} + \beta_{2}M_{\alpha,t} + \beta_{3}N_{i,t} + \mu_{t} + \varepsilon_{i,\alpha,t},
    \label{Eq:RegIndu}
\end{equation}
where $U_{i,\alpha,t+5}$ ($U_{i,\alpha,t}$) takes the value of 1 if $RCA_{i,\beta,t+5}\geq 1$ ($RCA_{i,\beta,t}\geq 1$) and 0 otherwise, $\omega_{i,\alpha,t}$ is the density of active related industries for industry $\alpha$ in province $i$ at year $t$, $M_{\alpha,t} = \sum_{i} U_{i,\alpha,t}$ is the number of provinces where that industry has revealed comparative advantage, $N_{i,t}= \sum_{\alpha} U_{i,\alpha,t}$ is the number of industries with revealed comparative advantage in that province, and $\varepsilon_{i,\alpha,t}$ is the error term. The regression equation includes the year-fixed effects, $\mu_{t}$, to control for any time-varying characteristics of provinces and industries.

The regression coefficient $\beta_{1}$ captures the impact of the density of active related industries in the probability of developing revealed comparative advantage in a new industry (see columns (1)-(3) of Table~\ref{Tab:RegIndu}) and in the probability of keeping revealed comparative advantage in an industry (see columns (4)-(6) of Table~\ref{Tab:RegIndu}). In all specifications we find the density of active related industries to be a strong, positive, and significant predictor of both, the probability of developing a new industry and keeping an industry in a Chinese province. In all cases, by controlling for the number of active industries in a province and the number of provinces that are active in an industry we show that our findings are not just a reflection of the industrial diversity of a province or the ubiquity of an industry.

\linespread{1}
\begin{table}[t]
  \centering
  \caption{Probit regressions for inter-industry learning.}
  \footnotesize
    \begin{tabular*}{\textwidth}{@{\extracolsep{\fill}}lcccccc}
    \toprule
    \multirow{3}[6]{*}{Independent Variables} & \multicolumn{6}{c}{Probit Model} \\
    \cmidrule{2-7}          & \multicolumn{3}{c}{Developing RCA in a Five-year Period} & \multicolumn{3}{c}{Keeping RCA in a Five-year Period} \\
    \cmidrule{2-7}          & (1)   & (2)   & (3)   & (4)   & (5)   & (6) \\
    \midrule
    \multirow{2}[1]{*}{Density of Active Related Industries} & \multicolumn{1}{l}{3.8844***} & \multicolumn{1}{l}{4.2084***} & \multicolumn{1}{l}{11.510***} & \multicolumn{1}{l}{-0.5753**} & \multicolumn{1}{l}{-1.7392***} & \multicolumn{1}{l}{15.266***} \\
          & \multicolumn{1}{l}{(0.1622)} & \multicolumn{1}{l}{(0.1661)} & \multicolumn{1}{l}{(0.3826)} & \multicolumn{1}{l}{(0.2266)} & \multicolumn{1}{l}{(0.2665)} & \multicolumn{1}{l}{(1.1658)} \\
    \multirow{2}[0]{*}{Number of Active Provinces in Industry} &       & \multicolumn{1}{l}{0.0559***} & \multicolumn{1}{l}{0.0624***} &       & \multicolumn{1}{l}{-0.0740***} & \multicolumn{1}{l}{-0.0834***} \\
          &       & \multicolumn{1}{l}{(0.0028)} & \multicolumn{1}{l}{(0.0029)} &       & \multicolumn{1}{l}{(0.0059)} & \multicolumn{1}{l}{(0.0062)} \\
    \multirow{2}[1]{*}{Number of Active Industries in Province} &       &       & \multicolumn{1}{l}{-0.1348***} &       &       & \multicolumn{1}{l}{-0.3101***} \\
          &       &       & \multicolumn{1}{l}{(0.0063)} &       &       & \multicolumn{1}{l}{(0.0193)} \\
    \midrule
    Observations & 25713 & 25713 & 25713 & 6837  & 6837  & 6837 \\
    Pseudo $R^2$ & 0.0626 & 0.0924 & 0.1217 & 0.0119 & 0.0497 & 0.1397 \\
    \bottomrule
    \end{tabular*}%
    \begin{flushleft}
    \emph{Notes}: Probit regressions modeling the probability of developing a new industry, or keeping an industry, in a Chinese province, as a function of the density of active related industries in a province, the number of provinces active in an industry, and the number of industries active in a province. Data is for the 2001-2015 period. Probit regressions include year-fixed effects. Significant level: $*p<0.1$, $**p<0.05$, and $***p<0.01$.
    \end{flushleft}
  \label{Tab:RegIndu}
\end{table}%

\subsection{Inter-regional learning}

Next we explore our data in search for evidence in support of inter-regional learning. Once again, we divide our analysis into three sections, a data visualization (for illustrative purposes), a graphical method, and a multivariate statistical model.

\linespread{1.5}
\begin{figure}[!t]
  \centering
  \includegraphics[width=0.99\textwidth]{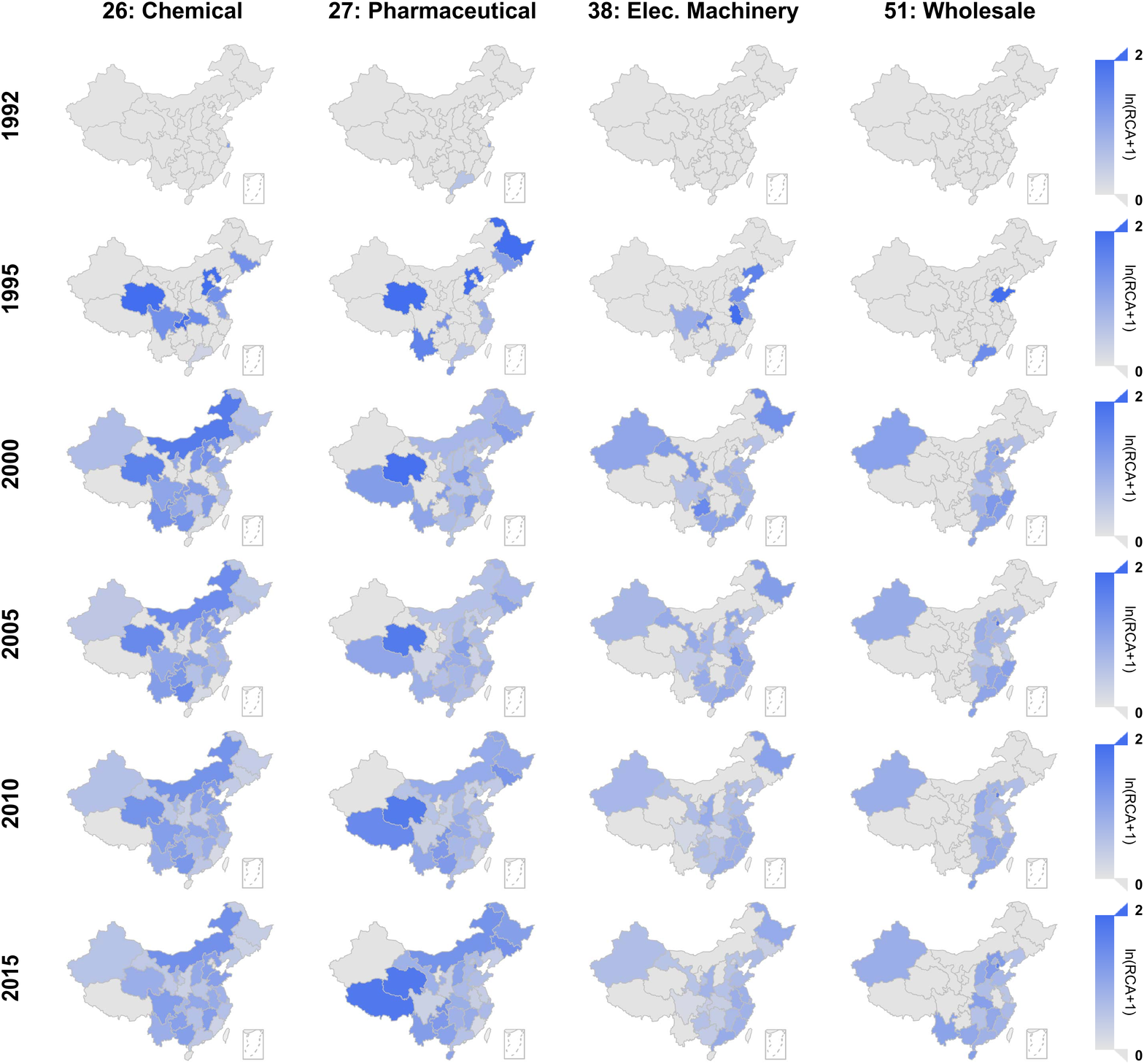}
  \caption{Evolution of revealed comparative advantage of provinces in China between 1992 and 2015. Four illustrated industries are Chemical Products Manufacturing Industry, Pharmaceutical Industry, Electric Machinery Manufacturing Industry, and Wholesale Industry (the keys of labels correspond to Figure~\ref{FigS2:CSRC}. The saturation of the color indicates the value of $ln(RCA+1)$.}
  \label{Fig4:ProvMap}
\end{figure}

Figure~\ref{Fig4:ProvMap} shows the spatial evolution of the presence of industries in Chinese provinces using data on the revealed comparative advantage of four industries (Chemical Products Manufacturing, Pharmaceuticals, Electric Machinery Manufacturing, and Wholesale) in each province between 1992 and 2015 (see Figure~\ref{FigS8:ProvMap} for an equivalent chart using the number of firms). The saturation of the color indicates the natural logarithm of the revealed comparative advantage of that province in that industry ($ln(RCA+1)$). In these four illustrative examples, we see that provinces that developed revealed comparative advantage in an industry tend to be neighbors of provinces that already had revealed comparative advantage in that industry, providing suggestive evidence for inter-regional learning.

Next, following \cite{Bahar2014}, we explore whether provinces in close physical proximity tend to have a more similar industrial structure. To do so, we measure the industrial similarity of a pair provinces using the cosine similarity of the vectors summarizing the revealed comparative advantage of industries in each province. Formally, let $y_{i,\alpha,t}=ln(RCA_{i,\alpha,t}+1)$ and $y_{j,\alpha,t}=ln(RCA_{j,\alpha,t}+1)$. Then, the industrial similarity $\varphi_{i,j,t}$ between provinces $i$ and $j$ at year $t$ will be given by:
\begin{equation}
  \varphi_{i,j,t}=\frac{ \sum_{\alpha}{ y_{i,\alpha,t} y_{j,\alpha,t} }}{\sqrt{\sum_{\alpha}{(y_{i,\alpha,t})^2}} \sqrt{\sum_{\alpha}{(y_{j,\alpha,t})^2}} }.
  \label{Eq:SimProv}
\end{equation}

\begin{figure}[t]
  \centering
  \includegraphics[width=0.9\textwidth]{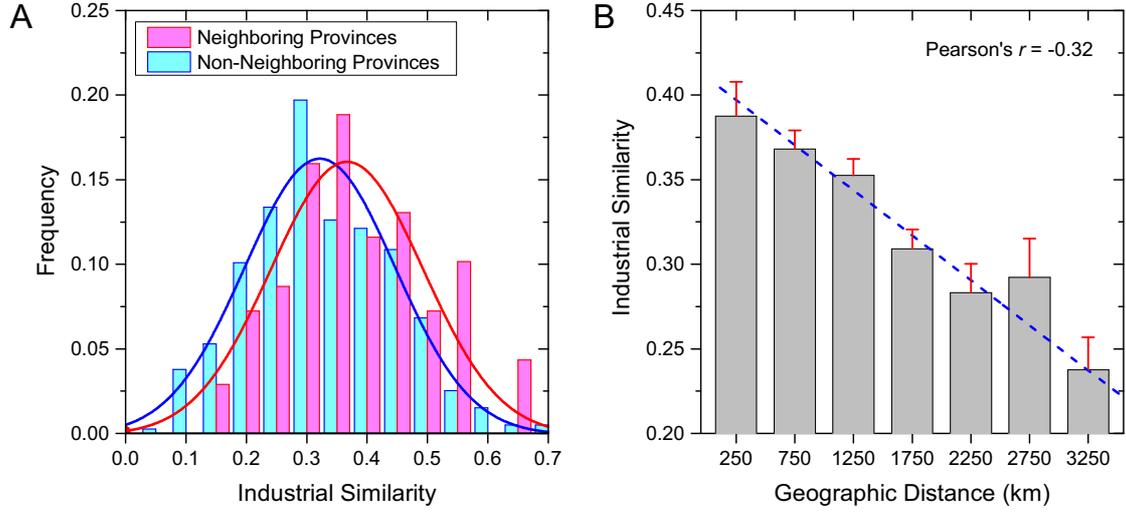}
  \caption{(A) Distribution of industrial similarity between pairs of neighboring provinces (in pink) and non-neighboring provinces (in blue). The red and blue curves are, respectively, normal fits for the distributions for neighboring and non-neighboring province pairs. (B) Industrial similarity between all pairs of provinces as a function of their geographic distance. Bars correspond to the average industrial similarity ($\varphi$) of pairs of provinces at that distance and error bars correspond to standard errors. The blue dash line represent a linear fit of the unbinned data. Pearson's correlation between industrial similarity and geographic distance is $r=-0.32$.}
  \label{Fig5:ProvNei}
\end{figure}

Figure~\ref{Fig5:ProvNei}A shows the distribution of the industrial similarities ($\varphi_{i,j}$) in 2015 for both, pairs of neighboring provinces (in pink) and pairs of non-neighboring provinces (in blue). We find that the industrial similarity of neighboring provinces is significantly larger than the similarity of non-neighboring provinces (ANOVA p-value=$8.1\times10^{-4}$). Figure~\ref{Fig5:ProvNei}B shows the industrial similarity ($\varphi_{i,j}$) as a function of geographic distance ($D_{i,j}$). Once again, we see that pairs of provinces in close physical proximity tend to be more similar than distant pairs of provinces (see Figure~\ref{FigS9:ProvCor} for equivalent charts using other distance and travel time measures).

Next, we formalize these observations by constructing an indicator, for each province, of the number of neighboring provinces that have developed revealed comparative advantage in each industry. We call this estimator the \emph{density} of active neighboring provinces ($\Omega$). For province $i$ in industry $\alpha$ at year $t$, the density of active neighboring provinces $\Omega_{i,\alpha,t}$ is given by:
\begin{equation}
    \Omega_{i,\alpha,t} = \left.{ \sum_{j} \frac{U_{j,\alpha,t}}{D_{i,j}} }\middle/
    \sum_{j} \frac{1}{D_{i,j}} \right.,
    \label{Eq:Provden}
\end{equation}
where $D_{i,j}$ is the geographic distance between provinces $i$ and $j$, and the binary variable $U_{j,\alpha,t}$ takes the value of 1 if $RCA_{j,\alpha,t}\geq 1$ and 0 otherwise.

Once again, we use the density estimator ($\Omega$) to explore whether the presence of a new industry in neighboring provinces increases the probability that this industry will appear in a province in the future. To perform this analysis, we estimate the density of active neighboring provinces ($\Omega$) for each province and industry in a base year and look at the new industries that appear in that province five years later. To reduce noise, we follow \cite{Bahar2014} and restrict the presence of new industries to two conditions: a backward condition, asking an industry to have an RCA below 1 for two years before the beginning of the period; and a forward condition, asking an industry to be present with RCA above 1 for two years after the end of the period.

\begin{figure}[t]
  \centering
  \includegraphics[width=0.9\textwidth]{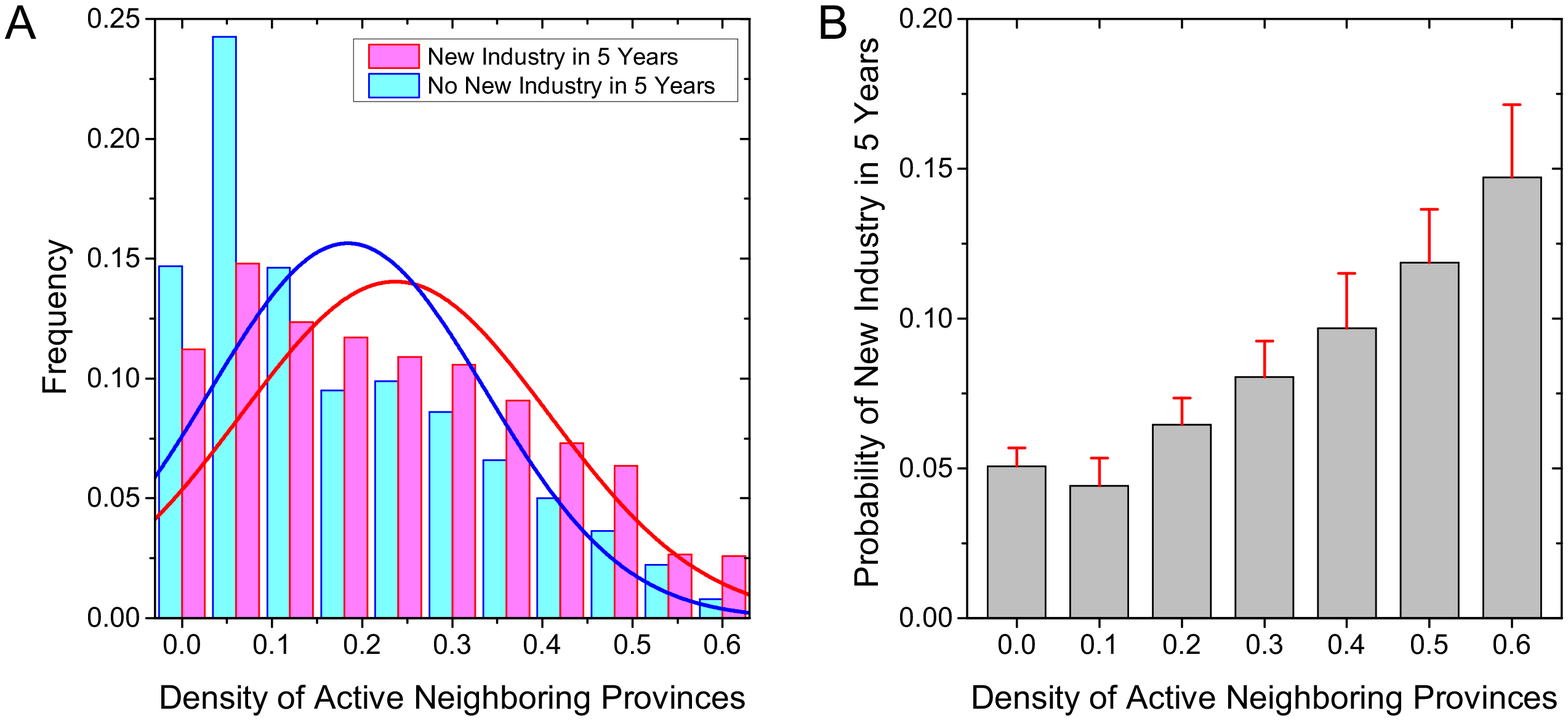}
  \caption{Inter-regional learning. (A) Distribution of the density of active neighboring provinces for each pair of provinces and industries. The pink distribution focuses only on pairs of provinces and industries that developed revealed comparative advantage in the next five years. The blue distribution is for the pairs of industries and provinces that did not develop revealed comparative advantage. The mean of the pink distribution is significantly larger than that of the blue distribution (ANOVA p-value=$1.4\times10^{-37}$). (B) Probability of a province developing comparative advantage in an industry as a function of the density of active neighboring provinces five years ago. Bars indicate average values and error bars indicate standard errors. Results show averages for 2001-2015 using five-year intervals.}
  \label{Fig6:ProvDen}
\end{figure}

Figure~\ref{Fig6:ProvDen}A compares the distribution of densities ($\Omega$) for industry-province pairs that developed revealed comparative advantage in an industry in a five-year period (in pink) and those that did not (in blue). We find that the average density of the province-industry pairs that developed revealed comparative advantage in a five-year period is significantly larger than the province-industry pairs that did not (ANOVA p-value=$1.4\times10^{-37}$).

Figure~\ref{Fig6:ProvDen}B shows the probability that a province will develop revealed comparative advantage in an industry as a function of the density of active neighboring provinces ($\Omega$). Once again, we find an increasing and convex relationship showing that the probability that a province will develop revealed comparative advantage in an industry increases strongly with the fraction of active neighboring provinces in that industry. These results are robust (see Figure~\ref{FigS10:ProvDen}) when we use other distance metrics in Eq.~(\ref{Eq:Provden}).

Finally, we use a multivariate probit model to estimate how the probability that a province will develop revealed comparative advantage, or keep revealed comparative advantage in an industry, is affected by the number of active neighboring provinces. We use this model to control for the number of industries in which that province already has revealed comparative advantage, and the number of provinces that already have comparative advantage in that industry. We estimate the following empirical specification:
\begin{equation}
    U_{i,\alpha,t+5} = \beta_{0} + \beta_{1}\Omega_{i,\alpha,t} + \beta_{2}N_{i,t} +\beta_{3}M_{\alpha,t} + \mu_{t} + \varepsilon_{i,\alpha,t},
    \label{Eq:RegProv}
\end{equation}
where $\Omega_{i,\alpha,t}$ is the density of active neighboring provinces for industry $\alpha$ and province $i$ at year $t$, and all other variables are defined as the same in Eq.~(\ref{Eq:RegIndu}).

\linespread{1}
\begin{table}[t]
  \centering
  \caption{Probit regressions for inter-regional learning.}
  \footnotesize
    \begin{tabular*}{\textwidth}{@{\extracolsep{\fill}}lcccccc}
    \toprule
    \multirow{3}[6]{*}{Independent Variables} & \multicolumn{6}{c}{Probit Model} \\
    \cmidrule{2-7}          & \multicolumn{3}{c}{Developing RCA in a Five-year Period} & \multicolumn{3}{c}{Keeping RCA in a Five-year Period} \\
    \cmidrule{2-7}          & (1)   & (2)   & (3)   & (4)   & (5)   & (6) \\
    \midrule
    \multirow{2}[1]{*}{Density of Active Neighboring Provinces} & \multicolumn{1}{l}{1.5393***} & \multicolumn{1}{l}{1.5621***} & \multicolumn{1}{l}{1.6969***} & \multicolumn{1}{l}{-1.4079***} & \multicolumn{1}{l}{-1.7160***} & \multicolumn{1}{l}{0.5836*} \\
          & \multicolumn{1}{l}{(0.0781)} & \multicolumn{1}{l}{(0.0782)} & \multicolumn{1}{l}{(0.2116)} & \multicolumn{1}{l}{(0.1317)} & \multicolumn{1}{l}{(0.1332)} & \multicolumn{1}{l}{(0.3311)} \\
    \multirow{2}[0]{*}{Number of Active Industries in Province} &       & \multicolumn{1}{l}{0.0404***} & \multicolumn{1}{l}{0.0402***} &       & \multicolumn{1}{l}{-0.0555***} & \multicolumn{1}{l}{-0.0660***} \\
          &       & \multicolumn{1}{l}{(0.0025)} & \multicolumn{1}{l}{(0.0025)} &       & \multicolumn{1}{l}{(0.0049)} & \multicolumn{1}{l}{(0.0051)} \\
    \multirow{2}[1]{*}{Number of Active Provinces in Industry} &       &       & \multicolumn{1}{l}{-0.0053} &       &       & \multicolumn{1}{l}{-0.1045***} \\
          &       &       & \multicolumn{1}{l}{(0.0075)} &       &       & \multicolumn{1}{l}{(0.0132)} \\
    \midrule
    Observations & 25713 & 25713 & 25713 & 6837  & 6837  & 6837 \\
    Pseudo $R^2$ & 0.0473 & 0.0648 & 0.0648 & 0.0342 & 0.0636 & 0.0793 \\
    \bottomrule
    \end{tabular*}
    \begin{flushleft}
    \emph{Notes}: Probit regressions modeling the probability of developing a new industry, or keeping an industry, in a Chinese province, as a function of the density of active neighboring provinces in an industry, the number of industries active in a province, and the number of provinces active in an industry. Data is for the 2001-2015 period. Probit regressions include year-fixed effects. Significant level: $*p<0.1$, $**p<0.05$, and $***p<0.01$.
    \end{flushleft}
  \label{Tab:RegProv}
\end{table}%

Table~\ref{Tab:RegProv} presents the results of our probit regressions. Once again, we divide our dataset into two sets: one containing all pairs of provinces and industries that do not have revealed comparative advantage (that we use to predict the ones that will develop RCA), and the other, with all province industry pairs with revealed comparative advantage (that we use to predict the ones who can sustain RCA).

Columns (1)-(3) of Table~\ref{Tab:RegProv} show the density of active related industries is a positive and significant predictor of the industries that a province will develop in the future, suggesting that provinces are more likely to develop an industry when they have neighbors that are competitive in that industry. The effect of active neighboring provinces on sustaining RCA in an industry, however, are not as clear (see columns (4)-(6) of Table~\ref{Tab:RegProv}). The bi-variate effects is negative, but becomes positive after controls. We interpret this as evidence of a tension between competition and learning, since an active neighboring province is a source of learning when that province does not have an industry, but it is also a source of competition when that province has developed that industry. In all cases, by controlling for the number of active industries in a province and the number of provinces that are active in an industry we show that our findings are not just a reflection of the industrial diversity of a province or the ubiquity of an industry.

\subsection{Combining inter-industry and inter-regional learning}

In the previous two sections we provided evidence supporting inter-regional and inter-industry learning in China's economic development. But do inter-regional and inter-industry learning work together? Or are they substitutes? In this section we combine both channels using graphical statistical methods and multivariate statistical models.

\linespread{1.5}
\begin{figure}[!t]
  \centering
  \includegraphics[width=0.6\textwidth]{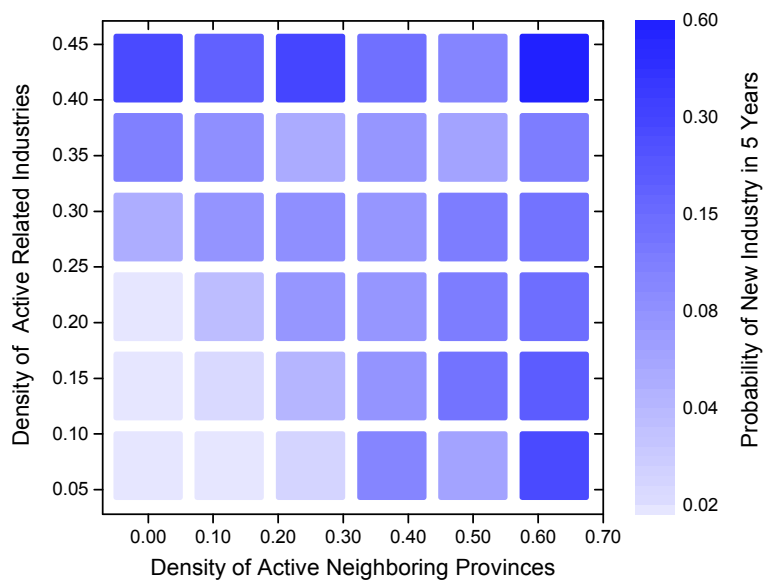}
  \caption{Joint probability of a province developing revealed comparative advantage in a new industry in a five-year period given the density of active neighboring provinces ($\Omega$) in horizontal-axis and the density of active related industries ($\omega$) in vertical-axis.}
  \label{Fig7:ProvIndu}
\end{figure}

First, we calculate the joint probability that a new industry will emerge in a province as a function of both the density of active neighboring provinces ($\Omega$) and the density of active related industries ($\omega$). All filters and definitions are equivalent to those used in the previous two sections. In agreement with our previous results, in Figure~\ref{Fig7:ProvIndu} we find that the probability that an industry will appear in a province in a five-year period increases with both, the density of active neighboring provinces ($\Omega$ at horizontal-axis) and the density of active related industries ($\omega$ at vertical-axis). The result is robust when using other density measures (see Figure~\ref{FigS11:ProvIndu}).

To explore the interaction between these two learning channels we use a probit model where the dependent variable $J_{i,\alpha,t+5}$ counts the number of provinces that developed comparative advantage in an industry. Once again we consider a backward and forward condition to reduce noise. Formally, $J_{i,\alpha,t+5}=1$ if $U_{i,\alpha,t}\: \&\: U_{i,\alpha,t-1} \: \& \: U_{i,\alpha,t-2} =0$ and $U_{i,\alpha,t+5}\: \&\: U_{i,\alpha,t+6}\: \&\:  U_{i,\alpha,t+7}=1$. The empirical specification is given by
\begin{equation}
    J_{i,\alpha,t+5} = \beta_{0} + \beta_{1}\Omega_{i,\alpha,t} + \beta_{2}\omega_{i,\alpha,t} + \beta_{3}\Omega_{i,\alpha,t}\omega_{i,\alpha,t} + \mu_{t} + \varepsilon_{i,\alpha,t},
    \label{Eq:ProvIndu}
\end{equation}
where $\Omega_{i,\alpha,t}$ is the density of active neighboring provinces, $\omega_{i,\alpha,t}$ is the density of active related industries, $\Omega_{i,\alpha,t}\omega_{i,\alpha,t}$ is the interaction term of the two densities, $\mu_{t}$ are the year-fixed effects, and $\varepsilon_{i,\alpha,t}$ is the error term.

Table~\ref{Tab:NPRI} presents the results of the probit regressions (see Table~\ref{Tab:StatJP} for summary statistics of regression variables). Column (1) shows the basic regression considering the density of active neighboring provinces ($\Omega$) and the density of active related industries ($\omega$). We find both effects are jointly significant. Column (2) adds an interaction term between the two densities ($\Omega \omega$). Once we add the interaction term we find the individual coefficients for both densities ($\Omega$ and $\omega$) to increase, while the interaction term is negative and significant. This indicates the presence of diminishing returns, meaning that the partial effect of each learning channel is reduced when the second channel is present. That is, when one learning channel is sufficiently active (inter-industry or inter-regional), the marginal contribution of the other one is reduced. Together, we find that the inter-industry learning has slightly stronger effect in activating new industries as suggested by its larger regression coefficient.

\linespread{1}
\begin{table}[!t]
  \centering
  \caption{Interaction between inter-industry learning and inter-regional learning.}
  \footnotesize
    \begin{tabular*}{\textwidth}{@{\extracolsep{\fill}}lcccccc}
    \toprule
    \multicolumn{1}{l}{\multirow{2}[4]{*}{New Industries in a Five-year Period}} & \multicolumn{6}{c}{Probit Model Using both Densities and Their Alternative Definitions} \\
    \cmidrule{2-7}          & (1)   & (2)   & (3)   & (4)   & (5)   & (6) \\
    \midrule
    \multirow{2}[1]{*}{Density of Active Neighboring Provinces} & \multicolumn{1}{l}{1.3092***} & \multicolumn{1}{l}{4.3405***} &       &       &       &  \\
          & \multicolumn{1}{l}{(0.0807)} & \multicolumn{1}{l}{(0.2421)} &       &       &       &  \\
    \multirow{2}[0]{*}{Density of Active Related Industries} & \multicolumn{1}{l}{3.7163***} & \multicolumn{1}{l}{ 6.2435***} &       &       &       &  \\
          & \multicolumn{1}{l}{(0.1713)} & \multicolumn{1}{l}{(0.2616)} &       &       &       &  \\
    \multirow{2}[0]{*}{Interaction Term 1} &       & \multicolumn{1}{l}{-11.8437***} &       &       &       &  \\
          &       & \multicolumn{1}{l}{(0.9136)} &       &       &       &  \\
    \multirow{2}[0]{*}{Ratio of Active Neighboring Provinces} &       &       & \multicolumn{1}{l}{0.5474***} & \multicolumn{1}{l}{0.6643***} &       &  \\
          &       &       & \multicolumn{1}{l}{(0.0499)} & \multicolumn{1}{l}{(0.0678)} &       &  \\
    \multirow{2}[0]{*}{Ratio of Active Related Industries} &       &       & \multicolumn{1}{l}{0.7802***} & \multicolumn{1}{l}{0.8502***} &       &  \\
          &       &       & \multicolumn{1}{l}{(0.0374)} & \multicolumn{1}{l}{(0.0472)} &       &  \\
    \multirow{2}[0]{*}{Interaction Term 2} &       &       &       & \multicolumn{1}{l}{-0.3701**} &       &  \\
          &       &       &       & \multicolumn{1}{l}{(0.1572)} &       &  \\
    \multirow{2}[0]{*}{Number of Active Neighboring Provinces} &       &       &       &       & \multicolumn{1}{l}{0.1739***} &  \\
          &       &       &       &       & \multicolumn{1}{l}{(0.0150)} &  \\
    \multirow{2}[0]{*}{Number of Active Related Industries} &       &       &       &       & \multicolumn{1}{l}{0.2093***} &  \\
          &       &       &       &       & \multicolumn{1}{l}{(0.0125)} &  \\
    \multirow{2}[0]{*}{Interaction Term 3} &       &       &       &       & \multicolumn{1}{l}{-0.0414***} &  \\
          &       &       &       &       & \multicolumn{1}{l}{(0.0065)} &  \\
    \multirow{2}[0]{*}{Number of Neighboring Provinces} &       &       &       &       &       & \multicolumn{1}{l}{0.0049} \\
          &       &       &       &       &       & \multicolumn{1}{l}{(0.0103)} \\
    \multirow{2}[0]{*}{Number of Related Industries} &       &       &       &       &       & \multicolumn{1}{l}{0.0178**} \\
          &       &       &       &       &       & \multicolumn{1}{l}{(0.0090)} \\
    \multirow{2}[1]{*}{Interaction Term 4} &       &       &       &       &       & \multicolumn{1}{l}{0.0010} \\
          &       &       &       &       &       & \multicolumn{1}{l}{(0.0019)} \\
    \midrule
    Observations & 25713 & 25713 & 25713 & 25713 & 25713 & 25713 \\
    Pseudo $R^2$ & 0.0819 & 0.0974 & 0.0653 & 0.0658 & 0.0658 & 0.0234 \\
    \bottomrule
    \end{tabular*}
    \begin{flushleft}
    \emph{Notes}: The regressions consider both effects of inter-regional learning and inter-industry learning. Data are for the 2001-2015 period. The probit regressions include the year-fixed effects. Significant level: $*p<0.1$, $**p<0.05$, and $***p<0.01$.
    \end{flushleft}
    \label{Tab:NPRI}
\end{table}

To check the robustness of our results we consider alternative definitions for both, the density of active neighboring provinces ($\Omega$) and the density of active related industries ($\omega$). In columns (3) and (4) of Table~\ref{Tab:NPRI} we repeat the exercise using simply the ratio of active neighboring provinces and the ratio of active related industries as independent variables (see online Appendix for details). This is equivalent to calculating both densities ($\Omega$ and $\omega$) using simple proportions instead of weighted averages. Once again, we find both effects are significant and there are diminishing returns to the addition of an alternative learning channel. Column (5) uses just the number of active neighboring provinces and the number of active related industries, instead of densities or ratios. We confirm the same results, although the explanatory power of this model is smaller than the one involving densities, meaning that the use of weighted averages to calculate densities contributes relevant information. Finally, column (6) presents a negative control: a model using the number of neighboring provinces and the number of related industries, no matter whether these are active or not. In this case, the model loses almost all its explanatory power and the effects are small, meaning that our results come from having active neighboring provinces and active related industries, but not from just having many neighboring provinces or just having many related industries.

\subsection{High-speed rail and inter-regional learning}

Finally, we study how the introduction of high-speed rail affected inter-regional learning using a differences-in-differences (DID) analysis. The introduction of high-speed rail is an adequate instrument because it reduces the barriers to inter-regional learning but should not affect inter-industry learning. In this section we check the effects of inter-regional learning first in terms of industrial similarity (measured by looking at the set of industries present in a province), and second, in terms of productivity (by looking at the increase in productivity of industries in provinces connected by high-speed rail)

During China's great economic expansion commercial train service was improved through several ``speed-up'' campaigns. These took the speed of trains from an average of only 48 km/h (in 1990s) to more than 300 km/h in the best cases \citep{Jiao2014}. By 2015, over 90 Chinese cities were connected by high-speed rail \citep{Lin2015}, and as of September 2016, China had the world's longest high-speed rail network, with over 20,000 km of track, a length that is longer than the rest of the world's high-speed rail tracks combined \citep{Cao2013}.

The introduction of high-speed rail reduced travel time among provinces, encouraging face-to-face interactions and potentially promoting learning among provinces \citep{Zheng2013}. Face-to-face interactions are considered important for learning, since they are a significant and effective way to build trust and to share complex ideas, even in the era of online communication technologies \citep{Storper2004}. The introduction of transport, or reductions in transportation costs, has been used in the past as instruments to test the effect of cost on the social interactions. For instance, \cite{Catalini2016} used the introduction of Southwest airlines, a discount airline in the US, to test whether reductions in ticket prices of direct flights between U.S. cities increased collaboration among scholars from the universities connected by these cheaper flights.

Similarly to \cite{Catalini2016} we address endogeneity concerns using the differences-in-differences (DID) method and the introduction of high-speed rail as an instrument. Because the introduction of high speed rail does not affect the similarity and productivity of industries within a province, this instrument help us isolate the effects of inter-regional learning from inter-industry learning.

The DID method requires two groups: treatment and control. In our DID analysis, pairs of provinces belong to the treatment group if they are connected by high-speed rail in 2015, otherwise they belong to the control group. Although there are many rounds of ``speed-up'' campaigns, we consider only the period between 2004 and 2014, since this allows us to capture the construction of numerous railroads (and hence obtain a larger sample), instead of observing just a few. The introduction of high-speed rail was identified using the Google Maps API in 2015 considering the accessibility between capital cities of provinces through high-speed rail passenger trains (see online Appendix for details).

We justify the use of the DID method as an identification strategy using two observations. On the one hand, the construction of high-speed rail between provinces should be close to random \citep{Qin2016}, at least with respect to the dependent variable (industrial similarity and productivity in our case and productivity of firms), and with respect to province level characteristics such as the levels of economic development and urbanization \citep{Bertrand2004,Besley2000}. This is because the construction of high speed rail was driven by political reasons (and not to connect provinces with similar industrial structures). For example, the ``Go West'' plan connected the coast with China's Far West. There is also the ``Silk Road Economic Belt'' plan \citep{Albalate2012,Rolland2015}, and plans to connect China with South East Asia \citep{Garver2006}. The construction of rail, therefore, can be seen as a quasi-experiment \cite{Catalini2016,Qin2016}. In fact, \cite{Qin2016} pointed out that the introduction of high-speed rail in China can be treated as a quasi-natural experiment because most new high-speed rails were implemented on existing railway lines instead of new railways.

\begin{figure}[!t]
  \centering
  \includegraphics[width=0.9\textwidth]{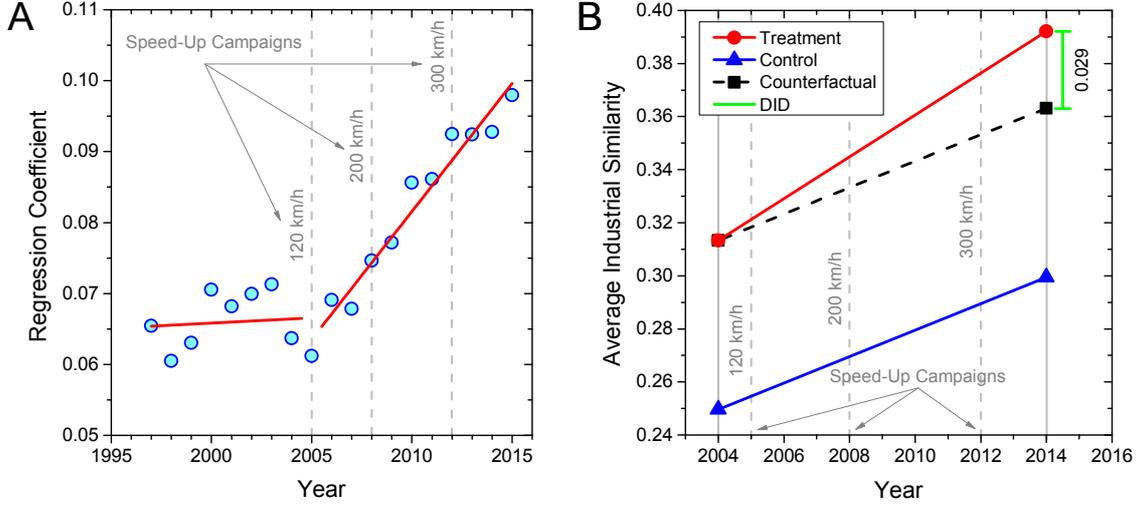}
  \caption{Industrial similarity and the introduction of high-speed rail. (A) Event study results. The $y$-axis shows the regression coefficient ($\beta_k$ in Eq.~(\ref{Eq:Event})) as a function of the year, after regressing the industrial similarity of pairs of provinces that were eventually connected by high-speed rail against the entry of high-speed rail. Red lines are linear fits for 1997-2005 and 2005-2015. (B) Differences-in-differences (DID) results. The $y$-axis is the average industrial similarity of all pairs of provinces connected by high-speed rail (in red) or not connected by high-speed rail (in blue). The value of DID (in green) is 0.029, and it is statistically significant. Vertical dash lines mark the years after speed-up campaigns, besides which the approximate average speeds of high-speed rail are shown.}
  \label{Fig8:HRail}
\end{figure}

On the other hand, the DID method is justified when the pre-trend of the dependent variable on the control and treatment groups is similar. Our data satisfies this condition prior to year 2005 (see Figure~\ref{Fig8:HRail}A for industrial similarity). To demonstrate this, we perform the event study by running the following ordinary least-squares (OLS) linear regression model using data between 1997 and 2015 in order to predict the industrial similarity between provinces $i$ and $j$ for each year as:
\begin{equation}
  \varphi_{i,j,t}= \beta_{0} + \sum_{k=1997}^{2015}\beta_{k}(Treat_{i,j}*1 \{t=k\}) + \varepsilon_{i,j}.
  \label{Eq:Event}
\end{equation}
Here $Treat_{i,j}$ is a dummy variable denoting whether provinces $i$ and $j$ are connected by high-speed rail and $1 \{t=k\}$ is an event time indicator, which is equal to 1 for the year where the pair was connected by high-speed rail. In other words, Eq.~(\ref{Eq:Event}) regresses the industrial similarity between pairs of provinces considering whether there is high-speed rail connecting them. Larger regression coefficients ($\beta_{k}$) tell us that the industrial similarity of the pairs of provinces connected by high-speed rail increased with respect to those that remain unconnected.

\linespread{1.5}

The results of Eq.~(\ref{Eq:Event}) are shown in Figure~\ref{Fig8:HRail}A. Before the introduction of high-speed rail (1997-2005), there is no temporal trend in $\beta_k$. After the introduction of high-speed rail (2005-2015), the effect of the treatment ($\beta_k$) begins to increase significantly, meaning that the treated provinces grew more similar after high-speed rail was introduced (see Figure~\ref{FigS12:NeigSim} for additional robustness check).

Next, we validate these results using differences-in-differences and the following specification:
\begin{equation}
  \varphi_{i,j,t}= \beta_{0} + \beta_{1}(Treat_{i,j}*After_{t}) + \beta_{2}Treat_{i,j} +\beta_{3}After_{t}+ \mathbf{A}\mathbf{X}^{\prime} +\varepsilon_{i,j}.
\label{Eq:DID}
\end{equation}
Here, $\varphi_{i,j,t}$ is the industrial similarity between provinces $i$ and $j$ at year $t$, and $\varepsilon_{i,j}$ is the error term. $Treat_{i,j}*After_{t}$ is the DID term, where the dummy $Treat_{i,j}$ denotes whether provinces $i$ and $j$ are affected by the introduction of high-speed rail. $After_{t}$ denotes whether it is before or after high-speed rail entry for each year $t$. The vector $\mathbf{X}$ denotes other control variables, which include gravity considerations, such as the difference between population, GDP per capita, urbanization, and trade, among province pairs.

\linespread{1}
\begin{table}[htbp]
  \centering
  \caption{DID regressions considering the effect of high-speed rail entry on the industrial similarity and the productivity of industries.}
  \footnotesize
    \begin{tabular*}{\textwidth}{@{\extracolsep{\fill}}lccccccc}
    \toprule
    \multirow{3}[6]{*}{Independent Variables} & \multicolumn{6}{c}{DID Regressions Using OLS Model} \\
\cmidrule{2-7}          & \multicolumn{3}{c}{Industrial Similarity} & \multicolumn{3}{c}{Productivity} \\
\cmidrule{2-7}          & (1)   & (2)   & (3)   & (4)   & (5)   & (6) \\
    \midrule
    \multirow{2}[1]{*}{High-speed Rail Entry} & \multicolumn{1}{l}{0.0290*} & \multicolumn{1}{l}{0.0266*} & \multicolumn{1}{l}{0.0268*} & \multicolumn{1}{l}{98713***} & \multicolumn{1}{l}{107343***} & \multicolumn{1}{l}{105636***} \\
          & \multicolumn{1}{l}{(0.0152)} & \multicolumn{1}{l}{(0.0150)} & \multicolumn{1}{l}{(0.0152)} & \multicolumn{1}{l}{(27649)} & \multicolumn{1}{l}{(27211)} & \multicolumn{1}{l}{(26011)} \\
    \multirow{2}[0]{*}{Treatment Group} & \multicolumn{1}{l}{0.0637***} & \multicolumn{1}{l}{0.0565***} & \multicolumn{1}{l}{0.0588***} & \multicolumn{1}{l}{39135**} & \multicolumn{1}{l}{30463*} & \multicolumn{1}{l}{26796} \\
          & \multicolumn{1}{l}{(0.0107)} & \multicolumn{1}{l}{(0.0110)} & \multicolumn{1}{l}{(0.0108)} & \multicolumn{1}{l}{(16240)} & \multicolumn{1}{l}{(17033)} & \multicolumn{1}{l}{(17379)} \\
    \multirow{2}[0]{*}{After Entry} & \multicolumn{1}{l}{0.0498***} & \multicolumn{1}{l}{0.0466***} & \multicolumn{1}{l}{0.0506***} & \multicolumn{1}{l}{364939***} & \multicolumn{1}{l}{376791***} & \multicolumn{1}{l}{361501***} \\
          & \multicolumn{1}{l}{(0.0091)} & \multicolumn{1}{l}{(0.0091)} & \multicolumn{1}{l}{(0.0090)} & \multicolumn{1}{l}{(17603)} & \multicolumn{1}{l}{(17362)} & \multicolumn{1}{l}{(16524)} \\
    \multirow{2}[0]{*}{$\Delta$ Population (log)} &       & \multicolumn{1}{l}{-0.0204***} &       &       & \multicolumn{1}{l}{-6881} &  \\
          &       & \multicolumn{1}{l}{(0.0049)} &       &       & \multicolumn{1}{l}{(8767)} &  \\
    \multirow{2}[0]{*}{$\Delta$ GDP per capita (log)} &       & \multicolumn{1}{l}{-0.0207**} &       &       & \multicolumn{1}{l}{109114***} &  \\
          &       & \multicolumn{1}{l}{(0.0081)} &       &       & \multicolumn{1}{l}{(17389)} &  \\
    \multirow{2}[0]{*}{$\Delta$ Urbanization} &       &       & \multicolumn{1}{l}{0.0160***} &       &       & \multicolumn{1}{l}{213686***} \\
          &       &       & \multicolumn{1}{l}{(0.0127)} &       &       & \multicolumn{1}{l}{(33900)} \\
    \multirow{2}[1]{*}{$\Delta$ Trade (log)} &       &       & \multicolumn{1}{l}{-0.0068***} &       &       & \multicolumn{1}{l}{20877***} \\
          &       &       & \multicolumn{1}{l}{(0.0024)} &       &       & \multicolumn{1}{l}{(4615)} \\
    \midrule
    Observations & 930   & 930   & 930   & 930   & 930   & 930 \\
    Robust $R^2$ & 0.1628 & 0.1833 & 0.1689 & 0.4980 & 0.5223 & 0.5548 \\
    RMSE  & 0.1109 & 0.1097 & 0.1106 & $2.10 \times 10^5$ & $2.00 \times 10^5$ & $2.00 \times 10^5$ \\
    \bottomrule
    \end{tabular*}
    \begin{flushleft}
    \emph{Notes}: Data are for the year 2004 (before high-speed rail entry) and 2014 (after high-speed rail entry). Significant level: $*p<0.1$, $**p<0.05$, and $***p<0.01$.
    \end{flushleft}
  \label{Tab:DID}
\end{table}

Figure~\ref{Fig8:HRail}B summarizes the results of the DID analysis studying the effect of high-speed rail on industrial similarity. The DID (in green) between treatment group (in red) and the expected trend from the control group (in dashed black line) is 0.029, indicating that pairs of provinces became more industrially similar after the introduction of high-speed rail. The first three columns of Table~\ref{Tab:DID} present the results of the DID regressions while controlling for differences in the level of population, GDP per capita, urbanization, and trade, among these pairs of cities (see Table~\ref{Tab:DStat} for summary statistics of covariates). The regression coefficient ($\beta_{1}$) of the interaction term ($Treat_{i,j}*After_{t}$) is positive and significant, and it is robust to controls (see Table~\ref{Tab:DID2}). These results suggest that the introduction of high-speed rail had an effect on the increase of industrial similarity experienced by pairs of Chinese provinces.

Second, we examine the effect of high-speed rail on inter-regional learning by measuring the productivity of industries. One may worry that the level of productivity is likely to rely on the industrial structure of provinces. However, the correlation coefficients between productivity and industrial similarity are neither high nor consistent over time, allowing us to explore the effect of high-speed rail on productivity as a separate observation (see Figure~\ref{FigS13:ProduCorr}).

Similar to what we did with before, we measure the \emph{productivity density} of active neighboring provinces as the average productivity of neighboring provinces weighted by distance. The productivity density of an industry in a province ($\zeta_{i\alpha}$) tells us if industry $\alpha$ in province $i$ is surrounded by provinces that are active and productive in that industry:
\begin{equation}
    \zeta_{i,\alpha,t} = \left.{ \sum_{j} \frac{\bar p_{i,j,\alpha,t}}{D_{i,j}} }\middle/
    \sum_{j} \frac{1}{D_{i,j}} \right.,
    \label{Eq:ProducDen}
\end{equation}
Here $\bar p_{i,j,\alpha,t}$ is the average productivity of provinces $i$ and $j$ in industry $\alpha$ at year $t$, and $D_{i,j}$ is the geographic distance between provinces $i$ and $j$. The productivity $\bar p$ of industry $\alpha$ in province $i$ is its labor productivity, measured as revenue per worker, i.e., the total revenue of industry $\alpha$ in a province $i$ divided by the total number of employees working in that industry $\alpha$ in that province $i$.

\begin{figure}[!t]
  \centering
  \includegraphics[width=0.9\textwidth]{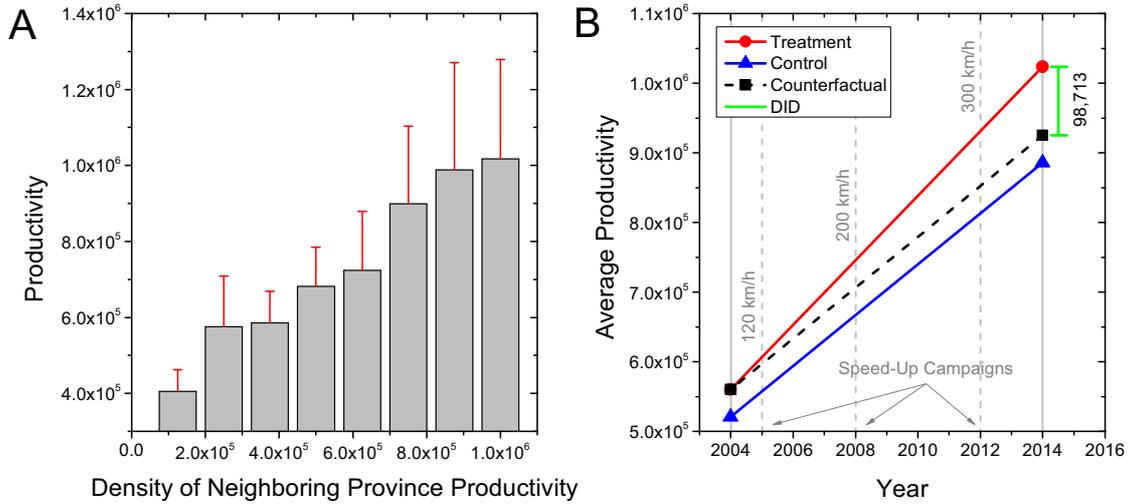}
  \caption{(A) Productivity of provinces as a function of the density of neighboring province productivity five years before. Bars indicate average values and error bars indicate standard errors. Results show averages for 2005-2014 using five-year intervals. (B) Average productivity of province pairs connected with high-speed rail (treatment, in red) and without high-speed rail (control, in blue). The differences-in-differences (DID, in green) is CNY 98,713 ($\sim$USD 15k). Vertical dash lines mark the years after speed-up campaigns, besides which the approximate average speeds of high-speed rail are shown.}
  \label{Fig9:Productivity}
\end{figure}

We use this density estimator ($\zeta$) to explore whether industries tend to be more productive when they are in provinces that are surrounded by neighbors that are productive in that industry. Figure~\ref{Fig9:Productivity}A shows that the average productivity of an industry in a province increases with the productivity density of neighboring provinces. Once again, we find an increasing and convex relationship.

\linespread{1.5}

Finally, we analyze the effects of high-speed rail on the average productivity of the industries in a province using differences-in-differences. Like before, we check the pre-trend of average productivity (see Figure~\ref{FigS14:EventP}A), and find there is no pre-trend (supporting the use of DID). For this DID analysis we modify Eq.~(\ref{Eq:DID}) by replacing the industrial similarity $\varphi_{i,j}$ with the average productivity $\bar p_{i,j}$ between pair of provinces $i$ and $j$. Figure~\ref{Fig9:Productivity}B shows a graphical summary of the DID analysis using average productivity. The DID (in green) between treatment group (in red) and control group (in blue) is CNY 98,713 ($\sim$USD 15k), meaning that workers in pairs of industries linked by high-speed rail increased their productivity, on average, by CNY 98,713 more than province pairs not connected by rail (see Figure~\ref{FigS14:EventP}B).

Finally, we present our differences-in-differences analysis for productivity and the instroduction of high-speed rail in the last three columns of Table~\ref{Tab:DID}. Here we find that the interaction term (High-speed Rail Entry) is positive and significant, and it is robust to controlling for differences in population, GDP per capita, urbanization, and trade. These results support the idea that the introduction of high-speed rail promoted learning, since the productivity of industries increased in the provinces that were connected by rail to provinces with productive firms in that industry.

\section{Conclusion and Discussion}

In this paper we explored the expansion of the Chinese economy between 1990 and 2015 by looking at the industrial diversification of Chinese provinces. First, we explored inter-industry learning by constructing the industry space, and showed that the probability that an industry will emerge in a province increased with the number of related industries already present in it. Next, we explored inter-regional learning and used geographic data to show that the probability that an industry will emerge in a province increases with the presence of the same industry in neighboring provinces. Then, we combined both of the results to study whether inter-industry and inter-regional learning reinforce each other, and found that the combination of the two learning channels exhibit diminishing returns, meaning that when one learning channel (inter-regional or inter-industry) is sufficiently active, the other channel does not contribute as much. That also implies that inter-regional learning and inter-industry learning are substitutes, or that learning is constrained by the absence of a single learning opportunity.

Moreover, we use the introduction of high-speed rail between provinces as an instrument to address endogenity concerns and provide evidence in support of the inter-regional learning hypothesis. We study how the introduction of high-speed rail affected the industrial similarity of the provinces connected by rail and the productivity of firms using differences-in-differences (DID). First, we show that the introduction of high-speed rail significantly increased the industrial similarity of the pairs of provinces connected by rail. Second, we compare the average productivity of industries that were present in pairs of neighboring provinces that became connected by rail, with that of pairs of neighboring provinces where these industries were present, but did not become connected by rail. We found that the average productivity of pairs of neighboring provinces that were connected by rail increased in the presence of a productive neighbor in that industry. These results provide evidence in support of inter-regional learning theories.

While encouraging, our results should be interpreted in the light of their limitations. For instance, the observed presence of new industries is limited to those with revealed comparative advantage in a province, instead of industries with a large absolute number of firms. That means industries without revealed comparative advantage are considered absent in our context, which can be a potential limitation. Also, our data and geographic resolution are limited. On the one hand, our data captures firms listed in China's two major stock markets (Shanghai and Shenzhen), which represent only a small fraction of all Chinese firms. Therefore, it is biased towards larger firms, since larger firms are more likely to be publicly listed. Moreover, some firms not listed or listed outside of China are not included even though they are located and operating in China. On the other hand, the use of provinces is also not ideal. Chinese provinces are relatively larger administrative units, some of which concentrate more than 100 million people. Improving the spatial resolution of this analysis would be an important improvement.

While we provide evidence in support of collective learning at the macro level, we do not provide a micro-channel for that learning. Is this learning the result of spin-off companies? Migrant workers? Supply and demand externalities? Labor market pooling? Or other channels? These micro level explanations are important, but escape the scope of this paper.

Nevertheless, the evidence presented here helps expand the body of literature supporting the idea that economic development is a learning rather than an accumulation process, and that learning is deeply path dependent, as it is affected by the presence of related industries and the industrial development of neighbors. This should be good news for developing countries looking to modernize their planning and economic development efforts. We hope this paper helps stimulate the study of collective learning in economic development, and also that it helps inspire new research to identify specific learning channels.

\section*{Acknowledgments}

We acknowledge the support from the MIT Media Lab Consortia and from the Masdar Institute of Technology. We also thank Haixing Dai, Mary Kaltenberg, Yiding Liu, Zhihai Rong, H. Eugene Stanley, Dan Yang, the Human Dynamics group meeting at the MIT Media Lab, and the B4 event by Research Center for Social Complexity (CICS) at Universidad del Desarrollo (UDD) for helpful comments. Jian Gao acknowledges the China Scholarship Council for partial financial support. This work was supported by the Center for Complex Engineering Systems (CCES) at King Abdulaziz City for Science and Technology (KACST) and the Massachusetts Institute of Technology (MIT).

\linespread{1.5}

%------------------------------------------------------------------------------%
%-----------------   Online  Appendix          --------------------------------%
%------------------------------------------------------------------------------%
\clearpage

\setcounter{figure}{0}
\setcounter{table}{0}
\setcounter{equation}{0}
\setcounter{page}{1}
\setcounter{section}{0}
\renewcommand{\thefigure}{S\arabic{figure}}
\renewcommand{\thetable}{S\arabic{table}}
\renewcommand\theequation{S\arabic{equation}}

\linespread{1}

\begin{center}
\Large{Collective Learning in China's Regional Economic Development}
\vspace{10pt}

\Large{(Online Appendix)}

\vspace{10pt}

\large{Jian Gao, Bogang Jun, Alex ``Sandy'' Pentland, Tao Zhou, C{\'e}sar A. Hidalgo}

\vspace{5pt}

\end{center}

\section{China's firm data}

\begin{figure}[!b]
  \centering
  \includegraphics[width=0.75\textwidth]{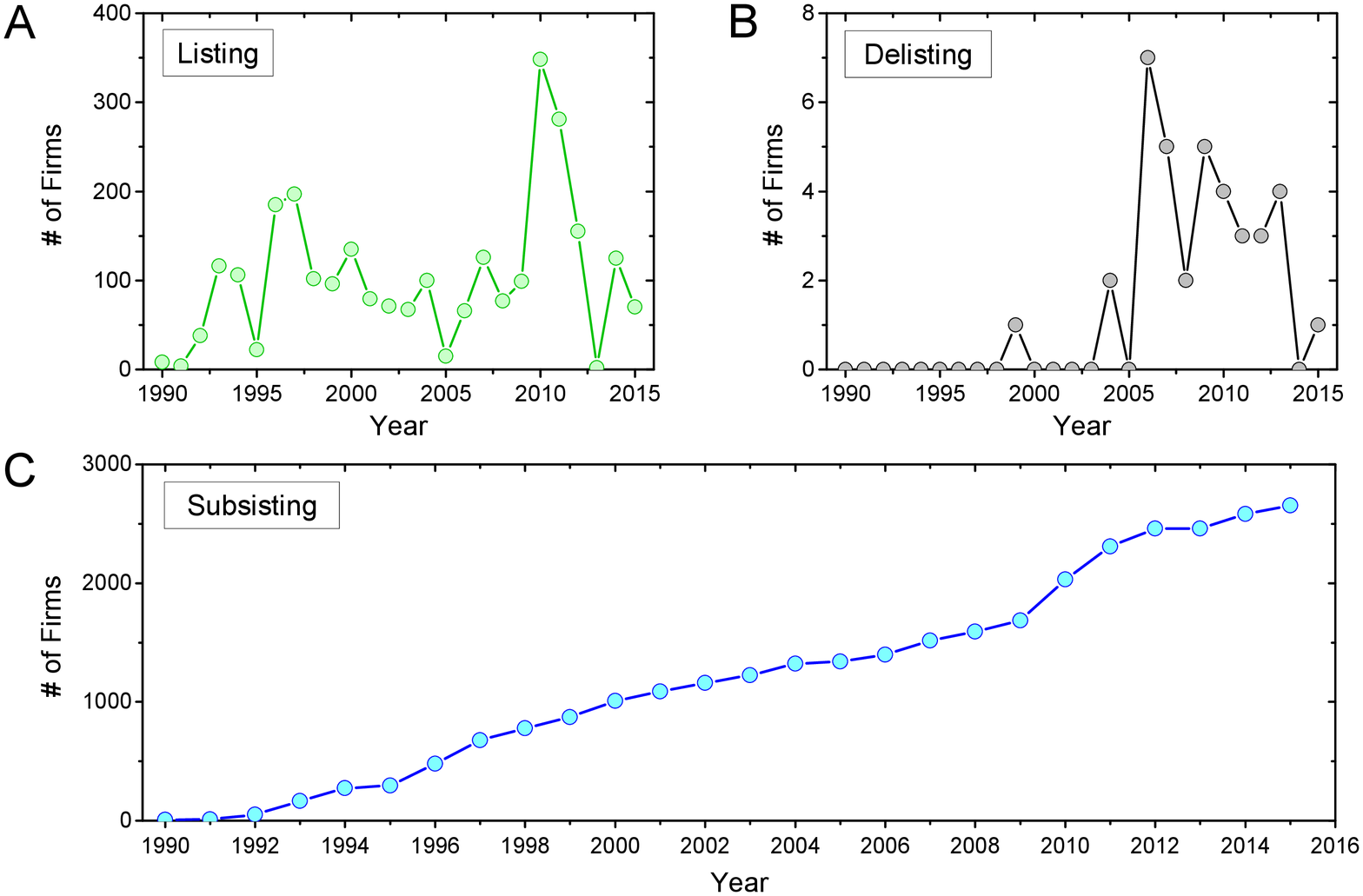}
  \caption{Number of firms which are listed, delisted and subsisting in each year. (A) Number of newly listed firms. (B) Number of newly delisted firms. (C) Number of subsisting firms, i.e., the cumulative number of listed firms which is not delisted yet.}
  \label{FigS1:Stat}
\end{figure}

\begin{figure}[!b]
  \centering
  \includegraphics[width=0.85\textwidth]{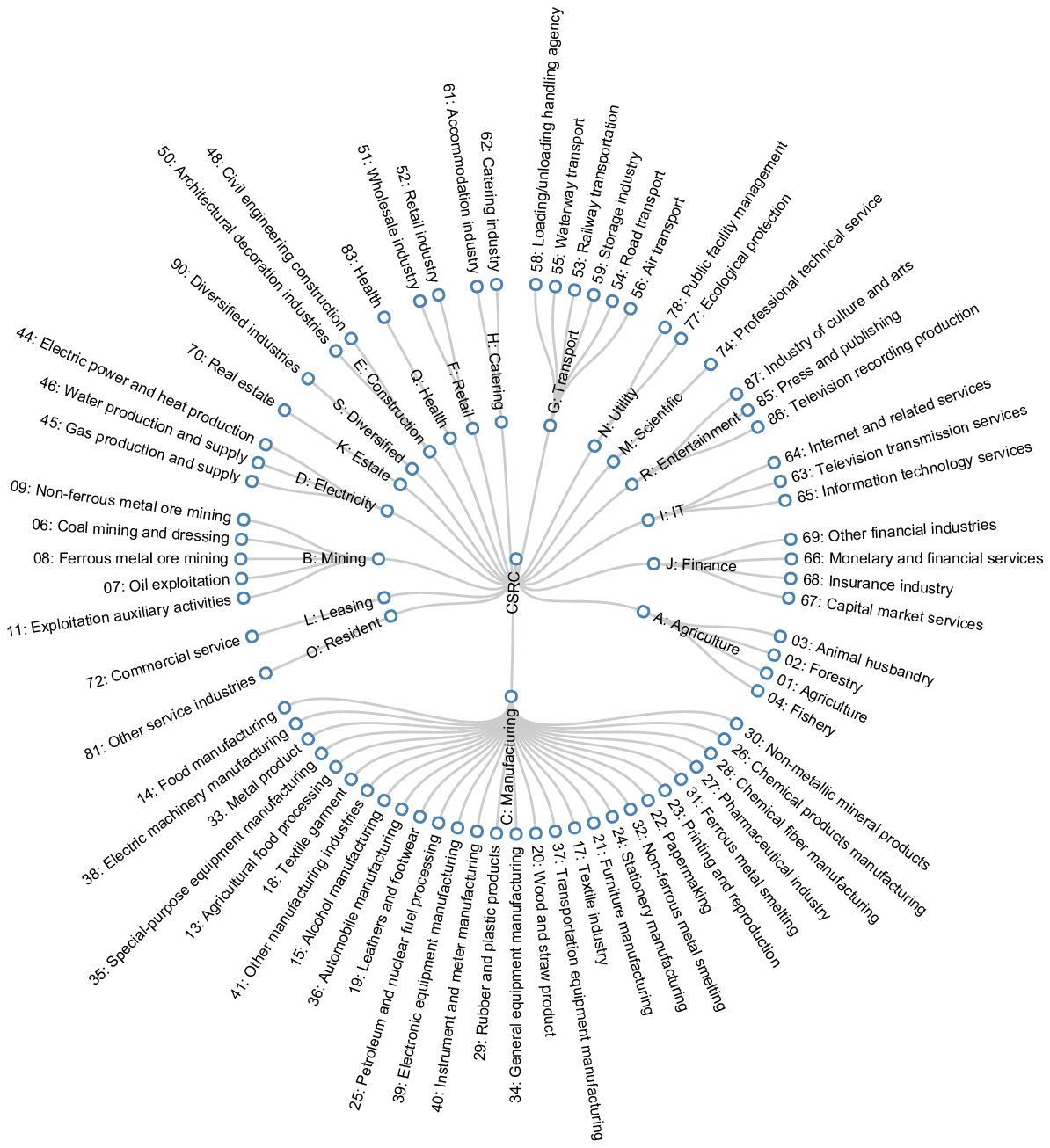}
  \caption{The codes of industry after the aggregation of firms into two levels: sectoral and sub-sectoral level. The inside layer corresponds to the sectoral level, while outside layer shows the sub-sectoral level. Here, we don't list all the sub-sectoral level.}
  \label{FigS2:CSRC}
\end{figure}

We use firm data from China's stock market extracted from the RESSET Financial Research Database, which is provided by Beijing Gildata RESSET Data Tech Co., Ltd. (http://www.resset.cn), a leading provider of economic and financial data in China. Our data set covers 1990-2015, a period during which China achieved rapid industrial development. This data set provides some basic registration information of publicly listed firms in Chinese stock exchanges, such as listing date, delisting date, registered address, industry category, yearly total revenue, and yearly number of employees. Although the numbers of newly listed and delisted firms in each year fluctuates, the overall number of firms increases almost linearly with time, as depicted in Figure~\ref{FigS1:Stat}. The registered addresses of firms cover 31 provinces in China. All these listed firms in our data set are aggregated into two levels, 18 categories and 70 subcategories. These categories are based on the ``Guidelines for the Industry Classification of Listed Companies'' issued by the China Securities Regulatory Commission (CSRC) (http://www.csrc.gov.cn) in 2011. CSRC category and CSRC subcategory codes as well as their associated industry names can be found in Figure~\ref{FigS2:CSRC}.

We aggregate those listed firm into two levels (see Figure~\ref{FigS2:CSRC}), which are sectoral and sub-sectoral level based on the ``Guidelines for the Industry Classification of Listed Companies'', which was issued by the China Securities Regulatory Commission (CSRC) in 2011. In this aggregation of firms into two levels, we consider the economic activities of listed firms with respect to the ``Industry Classification for the National Economy (GB T4754-2011)''. In the code of categories, which is depicted in Figure~\ref{FigS2:CSRC}, alphabet letters represent the sectoral level, while two-digit numbers represent the sub-sectoral level. To reduce noise, we only examine sub-sectoral level in which more than three firms are listed. After the aggregation, there are 2690 firms, as a total number of firms, that operate in 18 industries at sectoral level and 70 industries at sub-sectoral level.

\section{Distance, travel time, and macroeconomic indicators}

Regarding inter-regional learning, we use three distance metrics: geographic, driving, and neighboring distance (see  Table~\ref{Tab:Unit}). We define the geographic distance ($D_{i,j}$) between provinces $i$ and $j$ as a geodesic distance between the capital cities of the two provinces. The driving distance ($V_{i,j}$) is the shortest route between the capital cities of the two provinces, according to Google Maps API in 2015. The neighboring distance $B_{i,j}$ is defined as the least number of provinces that one province has to cross in order to reach another province. For example, the neighboring distance between Beijing and Shandong is two ($B_{i,j}=2$), because one has to cross at least two provinces to reach each other.

\begin{table}[!b]
  \centering
  \caption{Summary statistics of related economic indicators.}
    \footnotesize
    \begin{tabular*}{\textwidth}{@{\extracolsep{\fill}}llcccccc}
    \toprule
    \multicolumn{1}{c}{Variable} & \multicolumn{1}{c}{Description} & Unit  & Obs   & Min   & Max   & Mean  & Std. Dev. \\
    \midrule
    A. Province Level &       &       &       &       &       &       &  \\
    Population & Resident population at year-end & 10k person & 31    & $3.18\times 10^2$ & $1.07\times 10^4$ & $4.40\times 10^3$ & $2.80\times 10^3$ \\
    GDP per capita & Per capita gross domestic product & 1 CNY/person & 31    & $2.64\times 10^4$ & $1.05\times 10^5$ & $5.07\times 10^4$ & $2.21\times 10^4$ \\
    Urban Area & Total urban area in a region & 1 sq.km & 31    & $3.62\times 10^2$ & $2.13\times 10^4$ & $5.94\times 10^3$ & $5.17\times 10^3$ \\
    Land Area & Total land area in a region & 1 sq.km & 31    & $6.34\times 10^3$ & $1.66\times 10^6$ & $3.11\times 10^5$ & $3.87\times 10^5$ \\
    Trade & Total value of imports\&exports & 1k USD & 31    & $6.20\times 10^5$ & $1.24\times 10^9$ & $1.39\times 10^8$ & $2.51\times 10^8$ \\
    \midrule
    B. Province-pair Level &       &       &       &       &       &       &  \\
    Geographical Distance & Between two capital cities & 1k km & 465   & 114   & 3559  & 1369.4 & 723.0 \\
    Driving Distance & Between two capital cities & 1k km & 465   & 139   & 4883  & 1740.9 & 962.3 \\
    Neighboring Distance & Number of regions crossed & /     & 465   & 1     & 6     & 2.9   & 1.3 \\
    Transit Time & Shortest travel time by transit & 1 h   & 465   & 0.6   & 71    & 19.8  & 14.2 \\
    Normal-train Time & Shortest travel time normal-train & 1 h   & 465 &	1.6 &	71 &	25.5 &	14.3 \\
    Driving Time & Shortest travel time by driving & 1 h   & 465   & 1.9   & 59    & 19.4  & 11.5 \\
    $\Delta$ Population (log) & Difference in resident population & /     & 465   & 0.0071 & 3.5182 & 0.9330 & 0.7650 \\
    $\Delta$ GDP per capita (log) & Difference in GDP per capita & /     & 465   & 0.0000 & 1.3815 & 0.4502 & 0.3316 \\
    $\Delta$ Urbanization & Difference in urban area/land area & /     & 465   & 0.0053 & 262.31 & 7.4503 & 20.973 \\
    $\Delta$ Trade (log) & Difference in imports\&exports & /     & 465   & 0.0058 & 7.6024 & 1.9055 & 1.4401 \\
    \bottomrule
    \end{tabular*}
    \begin{flushleft}
    \emph{Notes}: The summary statistics of macroeconomic data, distance metrics and travel time measures are in 2014, 2015 and 2015, respectively.
    \end{flushleft}
  \label{Tab:Unit}
\end{table}

Regarding travel time, we consider three measures of it: transit, normal-train and driving time. The transit time is defined as the shortest time by high-speed trains. If there is no high-speed train on the whole route even by transfer, the shortest travel time by normal-train is used as an alternative. In this paper, the high-speed rail passenger trains refers to trains with code starting with ``G'', ``C'', and ``D'', while that of the normal-train staring with ``Z'', ``T'', ``K'' or using just numbers. The driving time is the shortest time when one travels between capital cities of the two provinces by drive. We estimate the driving time using the Google Maps API in 2015. The introduction of high-speed rail between two provinces was identified by using the Google Maps API in 2015 considering the accessibility between the two capital cities through high-speed rail passenger trains.

We collect macroeconomic data at the province-level, including Gross Domestic Product per capita (GDP per capita), population, total value of imports and exports, urban area, and total area (see Table~\ref{Tab:Unit}). The level of urbanization is defined as the share of urban area in a province. All of these macroeconomic indicators are from ``China's Statistical Yearbooks'', which are published by the National Bureau of Statistics of China (http://www.stats.gov.cn). These macroeconomic indicators cover the 1990-2015 period and 31 provinces.

Table~\ref{Tab:Unit} shows the brief descriptions and summary statistics of distance metrics, travel time measures, and macroeconomic indicators. At province level, we illustrate Population, GDP per capita, Urban Area, Land Area and Trade in 2014. At province-pair Level, we illustrate Geographical Distance, Driving Distance, Neighboring Distance, Transit Time, Normal-train Time and Driving Time in 2015, while $\Delta$ Population (log), $\Delta$ GDP per capita (log), $\Delta$ Urbanization, and $\Delta$ Trade (log) in 2014.

\section{Representation of industry space}

\begin{figure}[!t]
  \centering
  \includegraphics[width=0.8\textwidth]{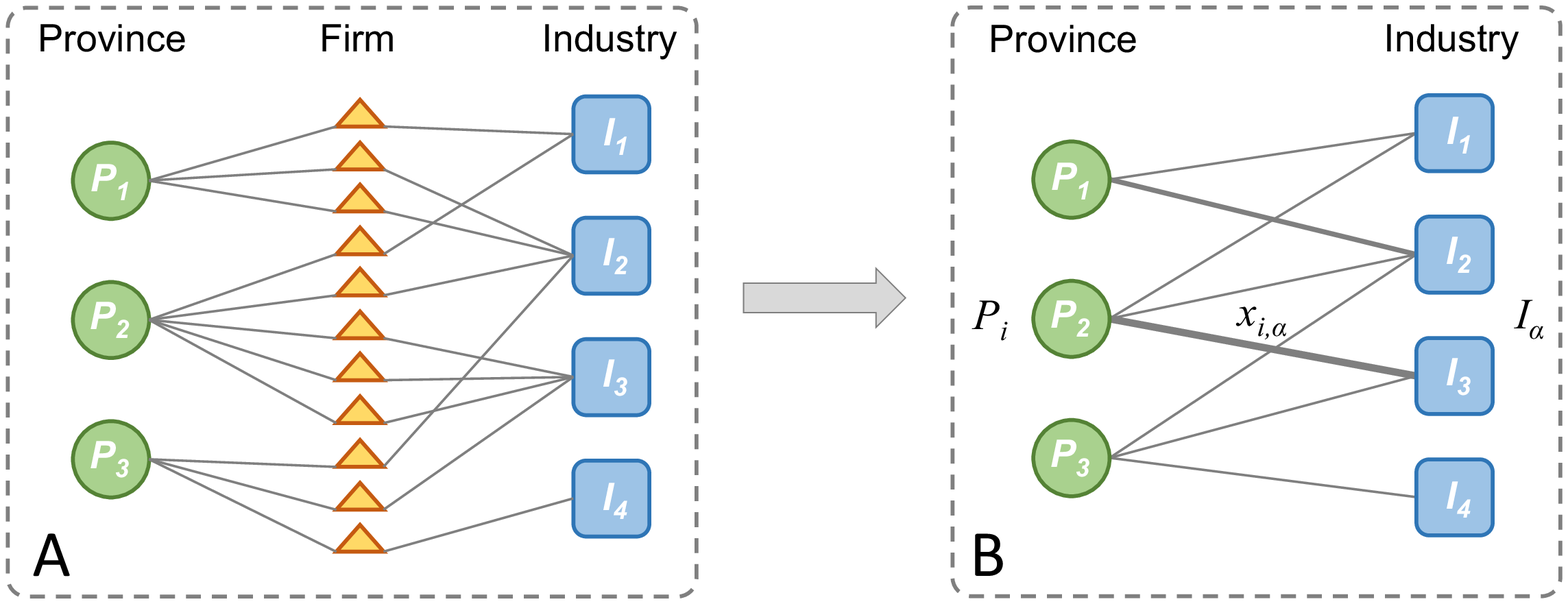}
  \caption{The bipartite network of ``province-industry'' pairs. $P$ and $I$ represent provinces and industries at the sub-sectoral level, respectively. The weight of link $x_{i,\alpha}$ corresponds to the number of firms in province $i$ that belong to industry $\alpha$.}
  \label{FigS3:BiNet}
\end{figure}

We build a ``province-industry'' bipartite network $G=\{P,I,E\}$ to connect provinces and industries (see Figure~\ref{FigS3:BiNet}), where $P$ is the set of provinces, $I$ is the set of industries at sub-sectoral level, and $E$ is the set of links. The weight of link $x_{i,\alpha}$ is the number of firms in province $i$ that operate in industry $\alpha$. In the following, $i$ and $\alpha$ indicate province-related and industry-related indices, respectively.

\begin{figure}[!t]
  \centering
  \includegraphics[width=0.90\textwidth]{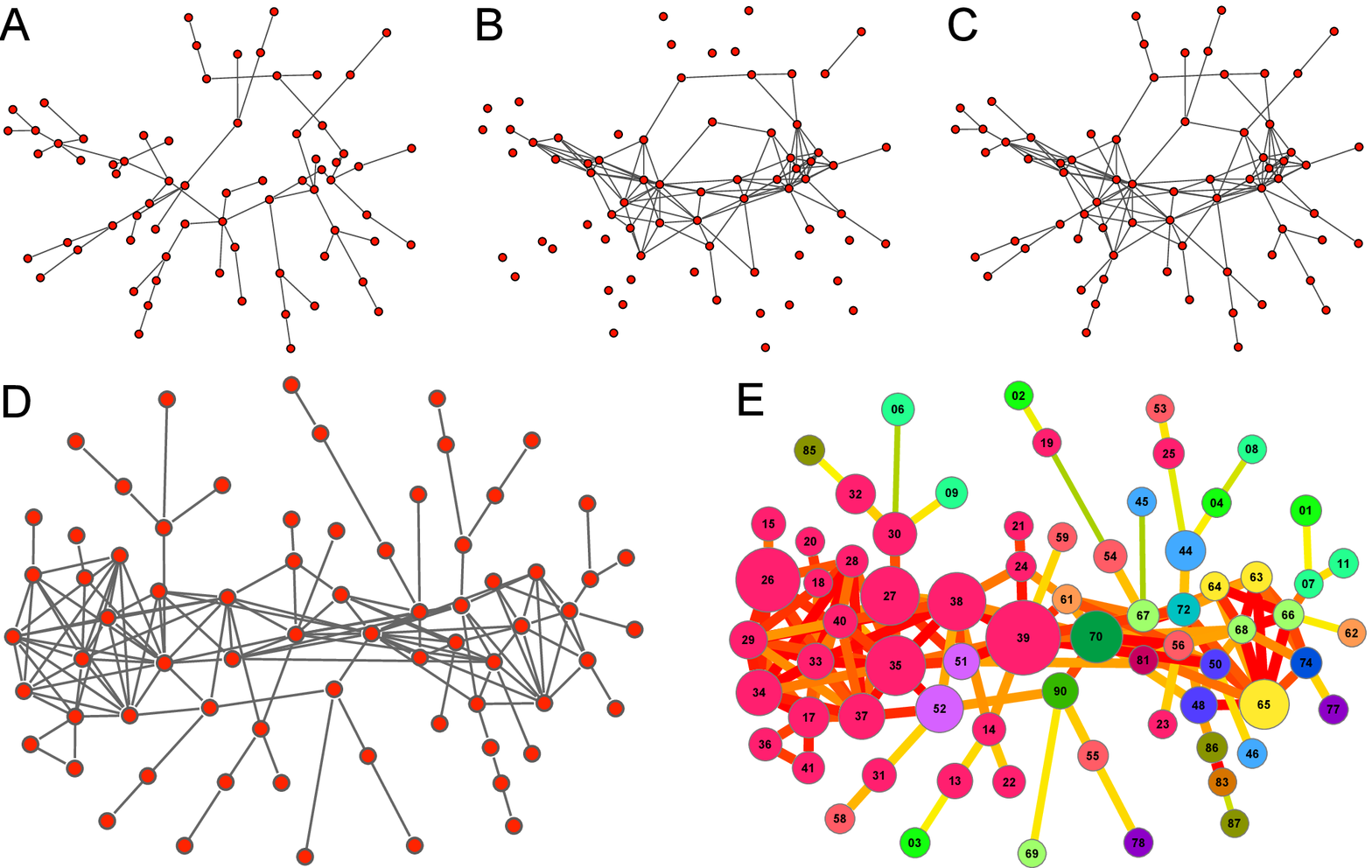}
  \caption{How to construct the industry space.(A) The first step: Building a maximum spanning network. (B) The second step: building a maximum weighted network with $\phi>0.81$. (C) The last step: Building a superposed network by combining the maximum spanning network and the maximum weighted network. (D) Layout of the product space, using a ForceAtlas2 algorithm in Gephi. (E) The final outcome: the industry space. The color of nodes corresponds to 18 industries at sectoral level. The size of nodes is proportional to the number of listed firms in that industry. The color and weight of links is associated with the $\phi$ value between two industries.}
  \label{FigS4:Net}
\end{figure}

To visualize the network of industry, we build a industry space using a proximity matrix $\Phi$, which is associated with the similarities between each pair of industries at the sub-sectoral level. There are three steps in building the industry space: (i) First step is to build a maximum spanning network, as shown in Figure~\ref{FigS4:Net}A. We calculate the maximum spanning tree so that all nodes becomes reachable in the network with minimum number of links. This network includes 69 links that ensures the connectivity and maximizes the total proximity. (ii) Then, we build a maximum weighted network, depicted in Figure~\ref{FigS4:Net}B, using only links of which weight exceeds a certain threshold $\phi'$. We set the threshold $\phi'$ as 0.81, under which the network includes 116 links and provides a distinguishable final visualization. (iii) Last, we combine these two networks, which are the maximum spanning network and the maximum wighted network, to build a superposed network (Figure~\ref{FigS4:Net}C). After the last step, the network includes 145 links and 70 nodes, which represent 70 industries at sub-sectoral level.

To make a better network visualization, we use a ForceAtlas2 algorithm of Gephi (http://gephi.github.io) in laying out the superposed network. ForceAtlas2 is a force directed layout, which places each node with consideration of the other nodes and allow to avoid overlapping links and untangle dense clusters. Figure~\ref{FigS4:Net}D shows the layout of industry space. After preparing the skeleton, we adjust the size of nodes according to the number of firms in that industry at the sub-sectoral level,and color each nodes according to the industries at the sectoral level. Likewise, we adjust the thickness and color of links according to the proximity. Finally, the industry space is depicted in Figure~\ref{FigS4:Net}E. The data of 2015 is used for the visualization of industry space.

\begin{figure}[t]
  \centering
  \includegraphics[width=0.9\textwidth]{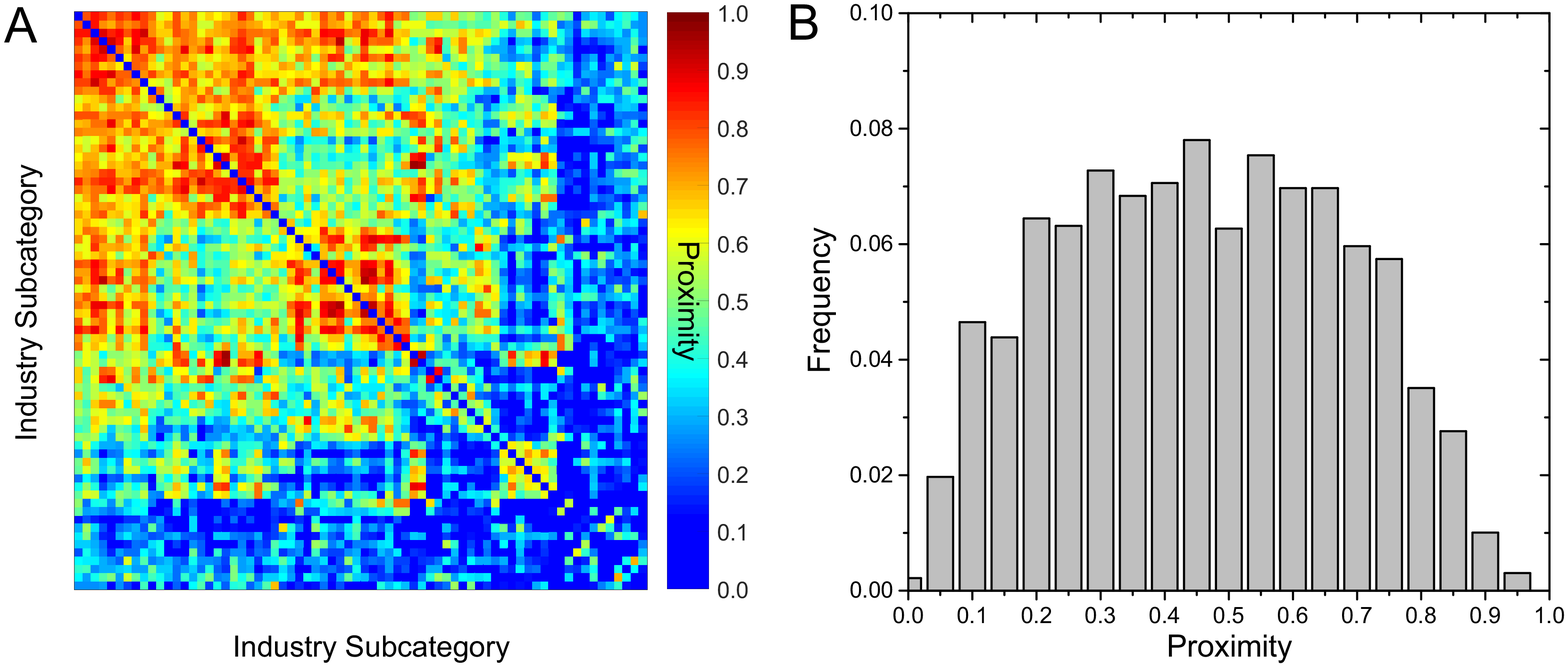}
  \caption{(A) Hierarchically clustered matrix based on the original proximity matrix ($\Phi$). The colors indicate the value of proximity. (B) The distribution of the proximity in matrix $\Phi$. The proximity matrix is calculated based on data in year 2015.}
  \label{FigS5:Modu}
\end{figure}

Regarding the proximity, Figure \ref{FigS5:Modu}A represents the proximity matrix $\Phi$ in a way of a hierarchically clustered matrix. The matrix shows two big modules and some small modules, supporting the existence of two density cores in industry space. Figure \ref{FigS5:Modu}B describes the density distribution of the proximity values in matrix $\Phi$. We can see that the vale of proximity follow a normal-like distribution with its average value around 0.5.

\section{Robustness check of inter-industry learning}

\begin{figure}[!t]
  \centering
  \includegraphics[width=0.9\textwidth]{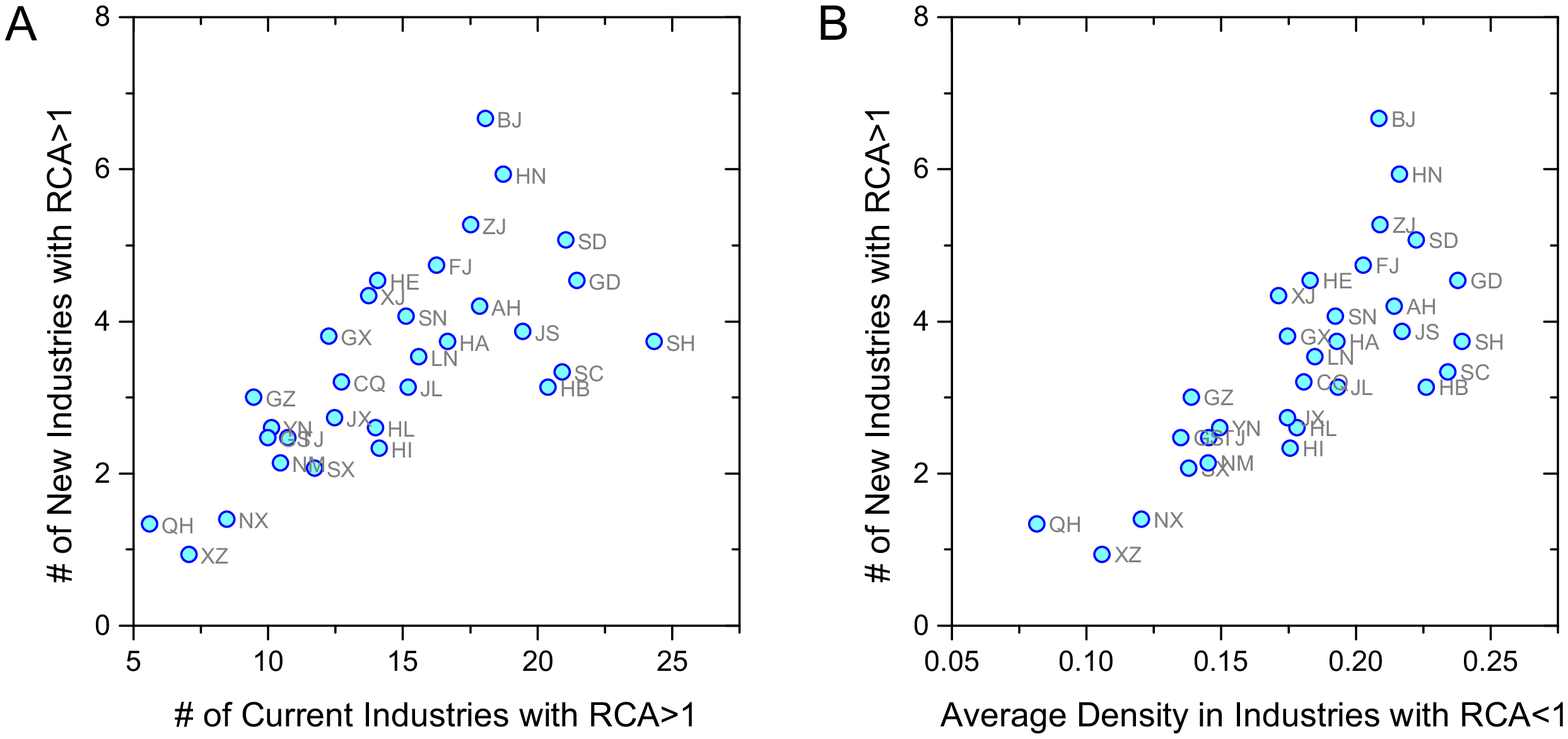}
  \caption{(A) Relationship between active industries at time $t$ and new active industries at time $t+5$ for provinces. (B) Relationship between the average density of active industries at time $t$ and new active industries at time $t+5$ for provinces. Results average for 2001-2015 using five-year intervals. Abbreviations of province names are shown in Table~\ref{Tab:Prov}.}
  \label{FigS6:InduNum}
\end{figure}

To check the robustness of inter-industry learning, here we also explore the relationship between the density of active related industries and the present of new industries in provinces. Figure~\ref{FigS6:InduNum}A presents the relationship between the number of industries, in which provinces have a revealed comparative advantage, and the number of new industries, in which provinces have developed a comparative advantage five years in the future. Using China's stock market data, we count the number of industries in year 2001 and check to see if new industries emerge five years in the future by looking at year 2006, and repeat this pattern over 2001 to 2015. More specifically, we will check the pairs of years (2002, 2007), (2003, 2008), ..., (2010, 2015).

In Figure~\ref{FigS6:InduNum}A, each dot represents each province of which value in horizontal-axis is corresponding to the average number of current industries with $RCA>1$ at year $t$ in time pairs in the period (2001-2015). The value in vertical-axis is corresponding to the average number of new industries with $RCA>1$ at time $t+5$ in time pairs, in which provinces didn't have a comparative advantage (i.e., with $RCA<1$) at the beginning ($t$) but developed a revealed comparative advantage (i.e., with $RCA>1$) five years later ($t+5$). There is a positive relationship between the number of industries in which a province has $RCA>1$ and the number of new industries in which the province diversified into five years later.

Figure~\ref{FigS6:InduNum}B presents the relationship between the average density in the industries with $RCA<1$ and the number of new industries in which a province develops a comparative advantage within five years. It shows strong positive relationship suggesting that the number of new industries is highly correlated with the density of related industries with $RCA<1$ five years before. In other words, the average density in industries with $RCA<1$ predicts the number of new industries that a province will diversify into in the future. See Table~\ref{Tab:Prov} for abbreviations of province names.

\begin{table}[!t]
  \centering
  \caption{Abbreviations of province names in China.}
  \footnotesize
    \begin{tabular*}{\textwidth}{@{\extracolsep{\fill}}cll|cll|cll}
    \toprule
    \multicolumn{1}{c}{ID}    & \multicolumn{1}{l}{Province Name} & Code   & \multicolumn{1}{c}{ID}    & \multicolumn{1}{l}{Province Name} & Code   & \multicolumn{1}{c}{ID}    & \multicolumn{1}{l}{Province Name} & Code \\
    \midrule
    1     & Beijing & BJ    & 12    & Anhui & AH    & 23    & Sichuan & SC \\
    2     & Tianjin & TJ    & 13    & Fujian & FJ    & 24    & Guizhou & GZ \\
    3     & Hebei & HE    & 14    & Jiangxi & JX    & 25    & Yunnan & YN \\
    4     & Shanxi & SX    & 15    & Shandong & SD    & 26    & Tibet & XZ \\
    5     & Inner Mongolia & NM    & 16    & Henan & HA    & 27    & Shaanxi & SN \\
    6     & Liaoning & LN    & 17    & Hubei & HB    & 28    & Gansu & GS \\
    7     & Jilin & JL    & 18    & Hunan & HN    & 29    & Qinghai & QH \\
    8     & Heilongjiang & HL    & 19    & Guangdong & GD    & 30    & Ningxia & NX \\
    9     & Shanghai & SH    & 20    & Guangxi & GX    & 31    & Xinjiang & XJ \\
    10    & Jiangsu & JS    & 21    & Hainan & HI    &       &       &  \\
    11    & Zhejiang & ZJ    & 22    & Chongqing & CQ    &       &       &  \\
    \bottomrule
    \end{tabular*}
  \label{Tab:Prov}
\end{table}

\begin{figure}[!t]
  \centering
  \includegraphics[width=0.9\textwidth]{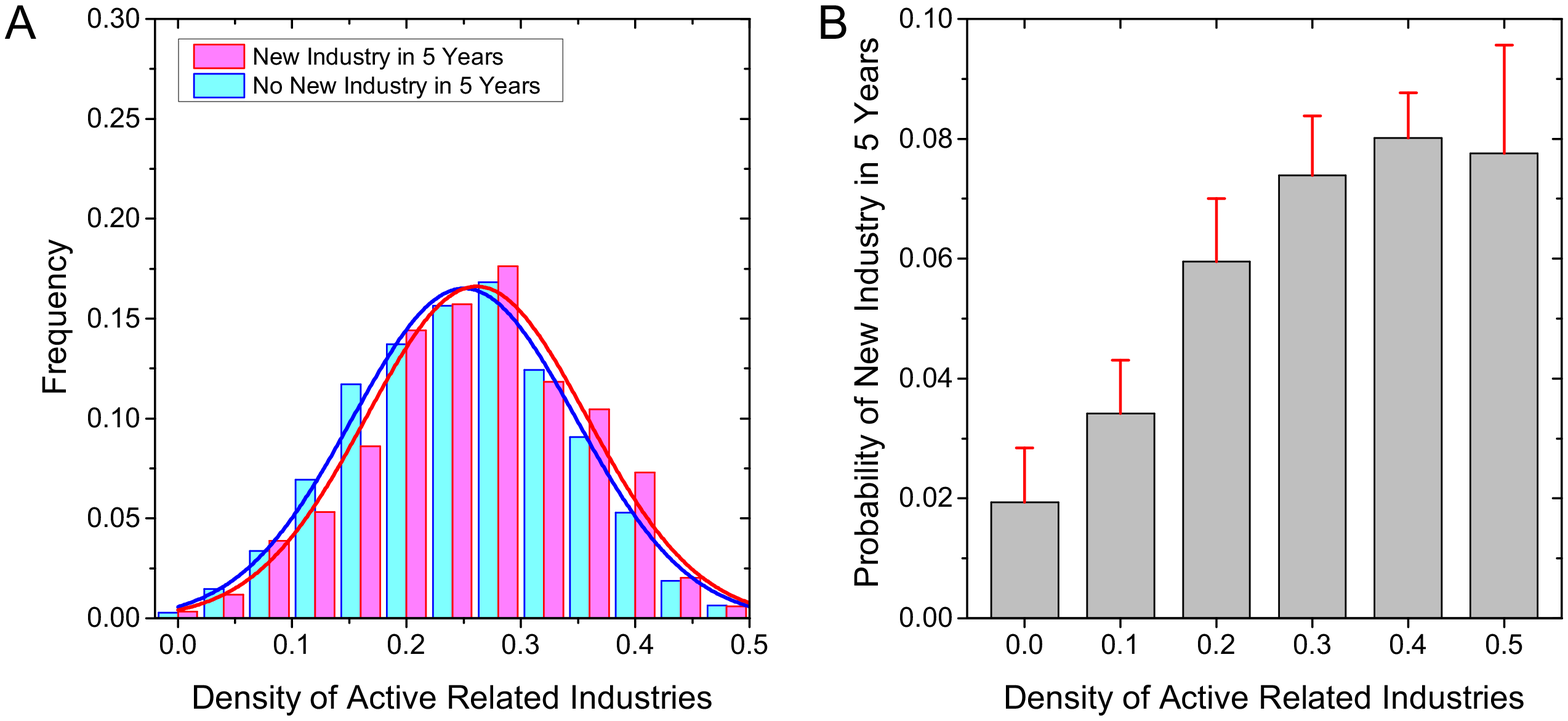}
  \caption{(A) Distribution of the density of active related industries for each pair of provinces and industries. The pink distribution focuses only on pairs of provinces and industries that developed revealed comparative advantage in the next five years. The blue distribution is for the pairs of industries and provinces that did not develop revealed comparative advantage. The mean of the pink distribution is significantly larger than that of the blue distribution (ANOVA p-value $<0.01$. (B) Probability that a new industry will appear in a province as a function of the density of active related industries ($\omega$). Bars indicate average values and error bars indicate standard errors. Results show averages for 2001-2015 using five-year intervals. The density $\omega$ in Eq.~\ref{Eq:Induden} in the main text is calculated using a time-varying industrial proximity ($\phi_{\alpha,\beta,t}$).}
  \label{FigS7:InduDen}
\end{figure}

To test the robustness of result seen in Figure~\ref{Fig3:InduDen} in the main text, we use a time-varying proximity ($\phi_{\alpha,\beta,t}$) in the main text when calculating the density of related industries in Eq.~(\ref{Eq:Induden}) in the main text. Again, we use China's stock market data with five years interval. Figure~\ref{FigS7:InduDen}A shows the distribution of related industry densities for pairs of industries and provinces that developed revealed comparative advantage (in pink) and that did not develop revealed comparative advantage (in blue) within five years. We find that the average related industry density for the pairs of industries and provinces in which developed revealed comparative advantage is significantly larger (ANOVA p-value $<0.01$). Figure~\ref{FigS7:InduDen}B shows an increasing and convex relationship meaning that the probability that an industry will develop revealed comparative advantage in a province increases strongly with the density of related industries that are already present in that province. These results provide robustness check of inter-industry learning.

\section{Robustness check of inter-regional learning}

\begin{figure}[!h]
  \centering
  \includegraphics[width=0.99\textwidth]{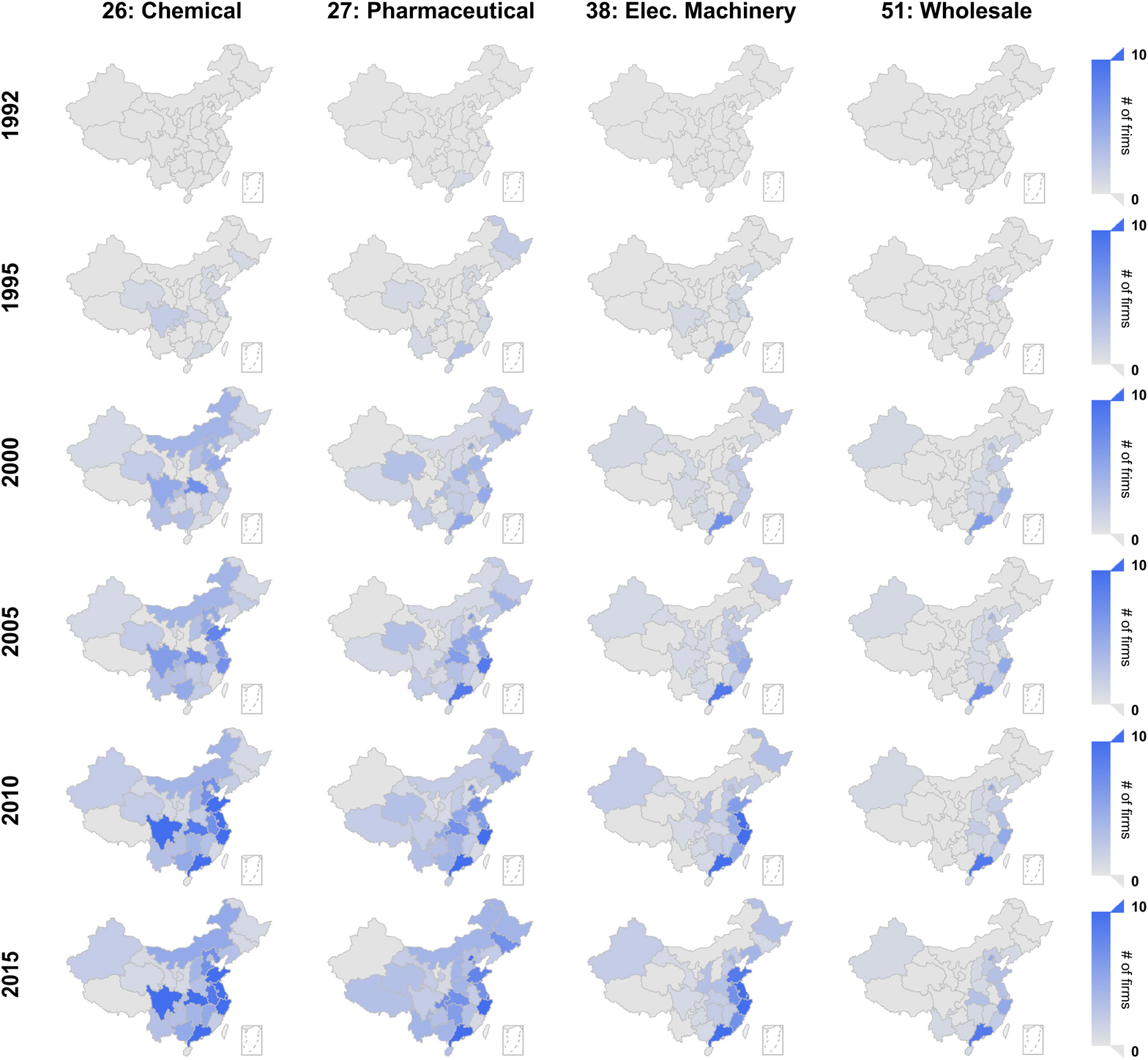}
  \caption{Evolution of the presence of industries in China between 1992 and 2015. Four illustrated industries are Chemical Products Manufacturing Industry, Pharmaceutical Industry, Electric Machinery Manufacturing Industry, and Wholesale Industry (the keys of labels correspond to Figure~\ref{FigS2:CSRC}. The saturation of the color indicates the number of firms.}
  \label{FigS8:ProvMap}
\end{figure}

To check the robustness of the observations in Figure~\ref{Fig4:ProvMap} in the main text, we additionally show the spatial evolution of the presence of industries in Chinese provinces using the number of firms of that industry in that province. Figure~\ref{FigS8:ProvMap}, for instance, presents the results of four industries: Chemical Products Manufacturing Industry, Pharmaceutical Industry, Electric Machinery Manufacturing Industry, and Wholesale Industry (the keys of labels correspond to Figure~\ref{FigS2:CSRC}. The saturation of the color indicates the number of firms. In this figure, the provinces that have large number of firms in an industry tend to be neighbors of provinces who already had a large number of firms in that industry, supporting our main finding.

\begin{figure}[!t]
  \centering
  \includegraphics[width=0.95\textwidth]{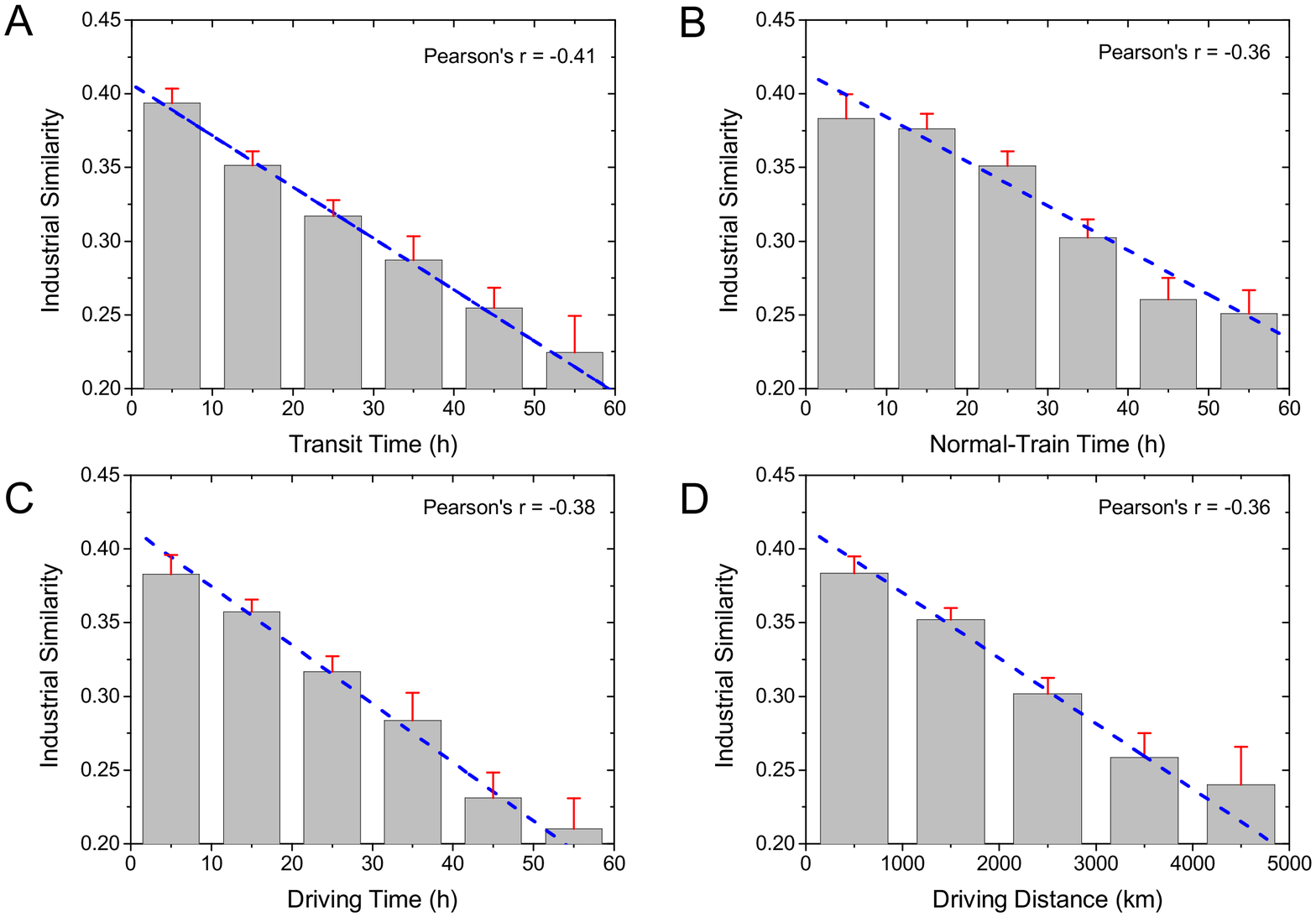}
  \caption{Relationship between industry similarity and (A) transit time, (B) normal-train time, (C) driving time, and (D) driving distance. Bar charts with error bars correspond to average values with stand errors in bins. Blue dash lines are linear fits of the corresponding bar charts.}
  \label{FigS9:ProvCor}
\end{figure}

Further, we present other evidences on the negative correlation between geographic proximity and the industrial similarity. More specific, in Figure~\ref{FigS9:ProvCor} we show the industrial similarity is highly correlated with transit time (A), normal-train time (B), driving time (C), and driving distance (D). We confirm that shorter travel time or closer distance between two regions corresponds to more similar industrial structure between the two.

To test the robustness of the results on inter-regional learning that is depicted in Figure~\ref{Fig6:ProvDen} in the main text, we develop an alternative index to measure the \emph{density} of active neighboring provinces. That is using neighboring distance $B_{i,j}$ to replace geographic distance $D_{i,j}$ when calculating the density of active neighboring provinces. Formally,
\begin{equation}
    \Omega_{i,\alpha,t} = \left.{ \sum_{j} \frac{U_{j,\alpha,t}}{B_{i,j}} }\middle/
    \sum_{j} \frac{1}{B_{i,j}} \right. .
    \label{EqS:Provden}
\end{equation}

\begin{figure}[!t]
  \centering
  \includegraphics[width=0.9\textwidth]{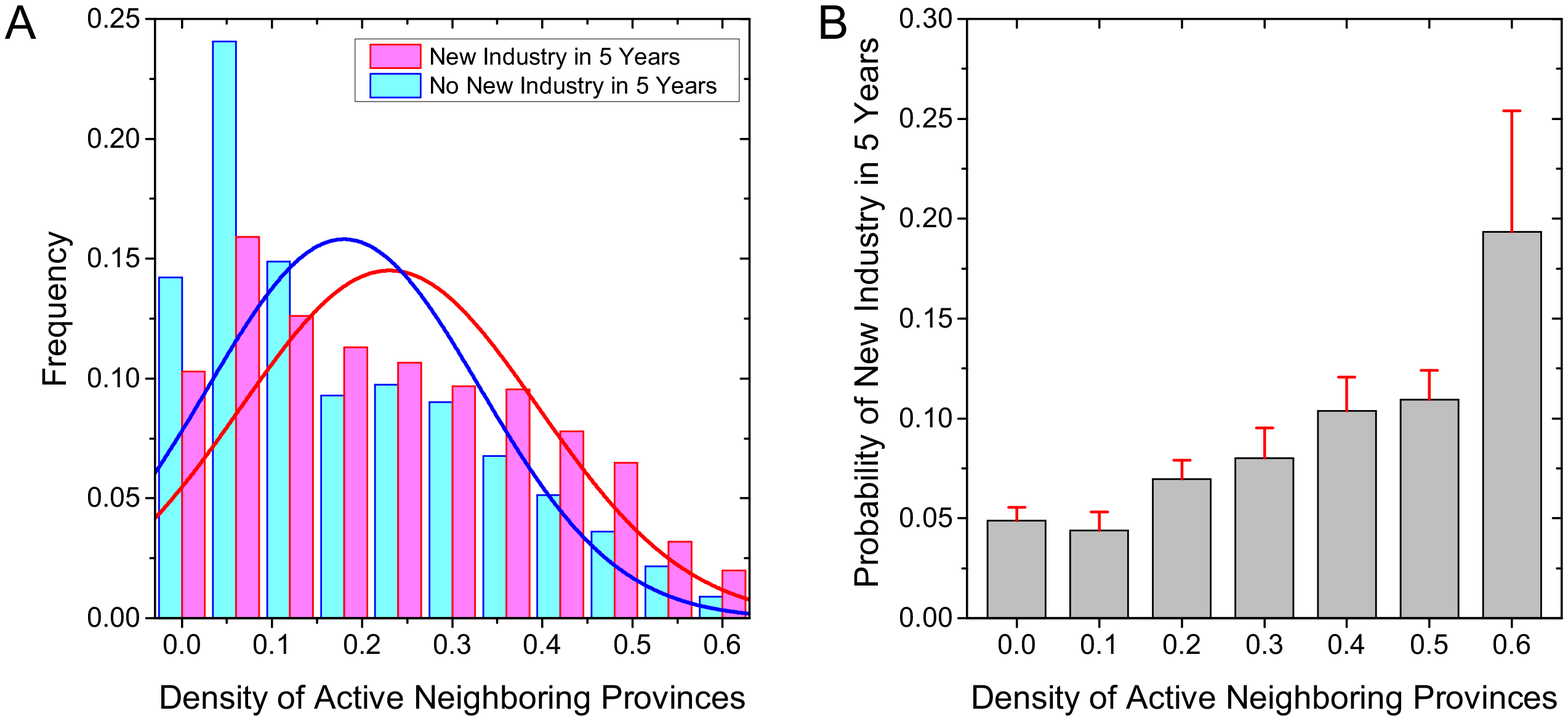}
  \caption{(A) Distribution of the density of active neighboring provinces for each pair of provinces and industries. The pink distribution focuses only on pairs of provinces and industries that developed revealed comparative advantage in the next five years. The blue distribution is for the pairs of industries and provinces that did not develop revealed comparative advantage. The mean of the pink distribution is significantly larger than that of the blue distribution (ANOVA p-value=$7.3\times10^{-36}$). (B) Probability of a province developing comparative advantage in an industry as a function of the density of active neighboring provinces five years ago. Bars indicate average values and error bars indicate standard errors. Results show averages for 2001-2015 using five-year intervals. The density $\Omega_{i,\alpha,t}$ in Eq.~(\ref{EqS:Provden}) is weighted by neighboring distance $B_{i,j}$.}
  \label{FigS10:ProvDen}
\end{figure}

Figure~\ref{FigS10:ProvDen}A shows the distribution of densities ($\Omega$) for industry-province pairs that developed revealed comparative advantage in an industry in a five-year period (in pink) and those who did not (in blue). We find that the average density of active neighboring province for the pairs of industries and provinces that developed revealed comparative advantage is significantly larger (ANOVA p-value=$7.3\times10^{-36}$). Figure~\ref{FigS10:ProvDen}B shows the increasing and convex relationship between the probability that a province will develop revealed comparative advantage in an industry and the density of active neighboring provinces in that industry five years before.

\section{Robustness check of inter-regional and inter-industrial learning}

\begin{figure}[!t]
  \centering
  \includegraphics[width=0.9\textwidth]{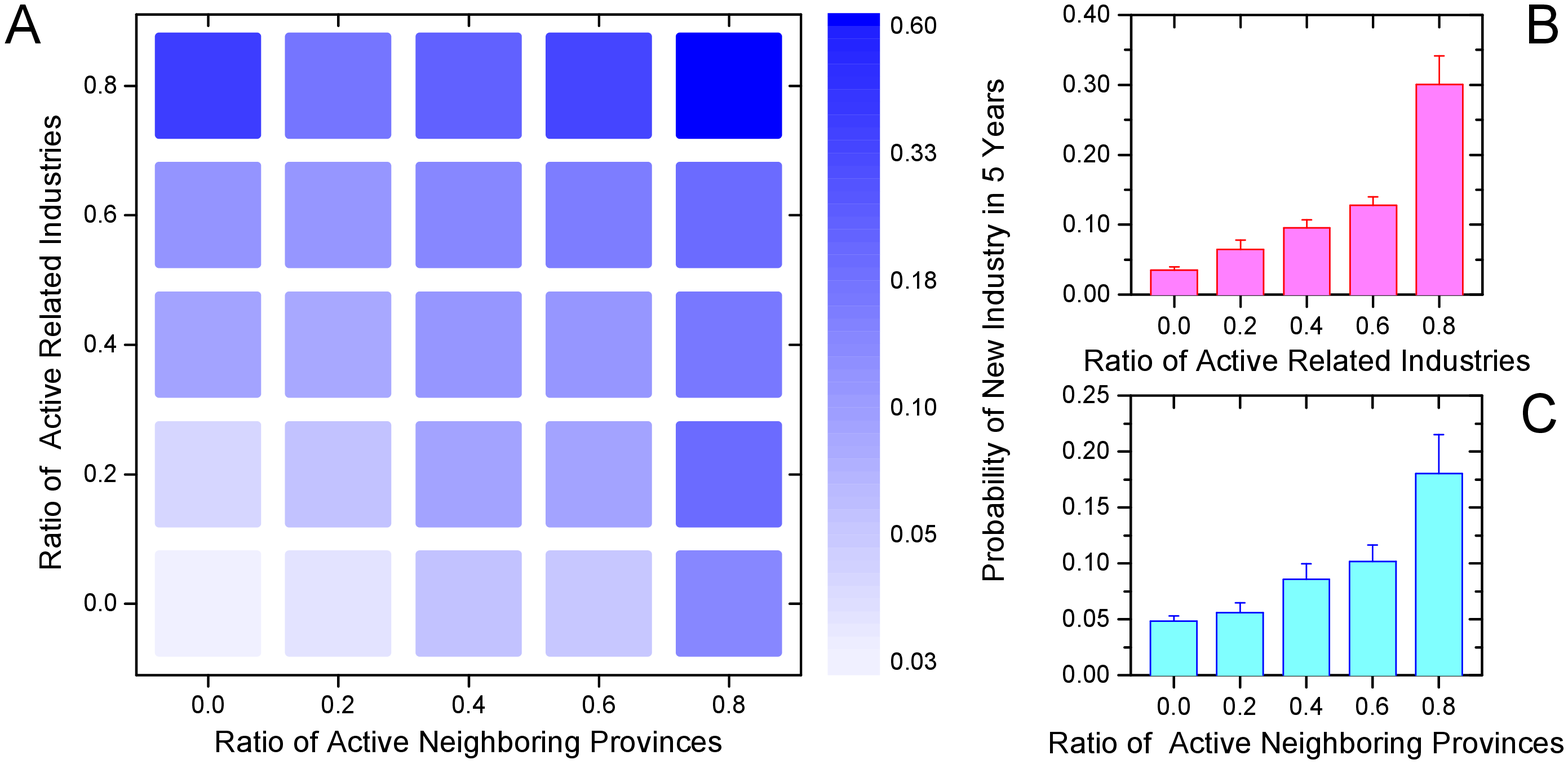}
  \caption{(A) Joint probability of a province developing revealed comparative advantage in a new industry in a five-year period, given the ratio of active neighboring provinces in horizontal-axis and the ratio of active related industries in vertical-axis. (B) and (C) are the corresponding marginal probability distributions of new industries present, given the ratio of active related industries and the ratio of active neighboring provinces, respectively.}
  \label{FigS11:ProvIndu}
\end{figure}

To check the robustness of our results in Figure~\ref{Fig7:ProvIndu} in the main text, we use an alternative index (ratio) to measure the density of active neighboring provinces ($\Omega$) and the density of related industries ($\omega$). For provinces, the ratio is the proportion of active neighboring provinces, and for industries, the ratio is the proportion of active related industries according to the illustrated industry space in 2015. Figure~\ref{FigS11:ProvIndu}A shows the joint probability of new industries present, given the ratio of active neighboring provinces and the ratio of active related industries. We can see that both ratios have significant effects on the new industries present. For each single effect, the ratio of active related industries (see Figure~\ref{FigS11:ProvIndu}B) and the ratio of active neighboring provinces (see Figure~\ref{FigS11:ProvIndu}C), the increasing and convex relationship shows that the probability that an industry will develop revealed comparative advantage in a province increases strongly with the ratio. These results support the robustness of our results.

Table~\ref{Tab:StatJP} presents the summary statistics of regression variables that are used in econometrics considering both inter-regional and inter-industry learning. Four different groups of metrics are included in the multivariable regressions: density, ratio, number, and active number. Here, the density of active related industries, the number of related industries, and the number of active related industries are all based on the illustrated industry space in 2015.

\begin{table}[!t]
  \centering
  \caption{Summary statistics of regression variables in the analysis of the emergence of new industries.}
  \footnotesize
    \begin{tabular*}{\textwidth}{@{\extracolsep{\fill}}lccccc}
    \toprule
    \multicolumn{1}{c}{Variable} & Observations   & Min   & Max   & Mean  & Std. Dev. \\
    \midrule
    Density of Active Neighboring Provinces & 25713 & 0     & 0.7127 & 0.1894 & 0.1551 \\
    Density of Active Related Industries & 25713 & 0.0109 & 0.5939 & 0.2283 & 0.0866 \\
    Interaction Term 1 & 25713 & 0     & 0.3022 & 0.0453 & 0.0441 \\
    Ratio of Active Neighboring Provinces & 25713 & 0     & 1     & 0.1816 & 0.2358 \\
    Ratio of Active Related Industries & 25713 & 0     & 1     & 0.1949 & 0.2884 \\
    Interaction Term 2 & 25713 & 0     & 1     & 0.0457 & 0.1063 \\
    Number of Active Neighboring Provinces & 25713 & 0     & 7     & 0.7935 & 1.0441 \\
    Number of Active Related Industries & 25713 & 0     & 9     & 0.8448 & 1.2864 \\
    Interaction Term 3 & 25713 & 0     & 45     & 1.0833 & 2.9363 \\
    Number of Neighboring Provinces & 25713 & 1     & 8     & 4.4463 & 1.8048 \\
    Number of Related Industries & 25713 & 1     & 15    & 3.8851 & 3.3691 \\
    Interaction Term 4 & 25713 & 1     & 120     & 17.271 & 17.712 \\
    \bottomrule
    \end{tabular*}
  \label{Tab:StatJP}
\end{table}

\section{Robustness check of causal evidence for inter-regional learning}

\begin{figure}[t]
  \centering
  \includegraphics[width=0.9\textwidth]{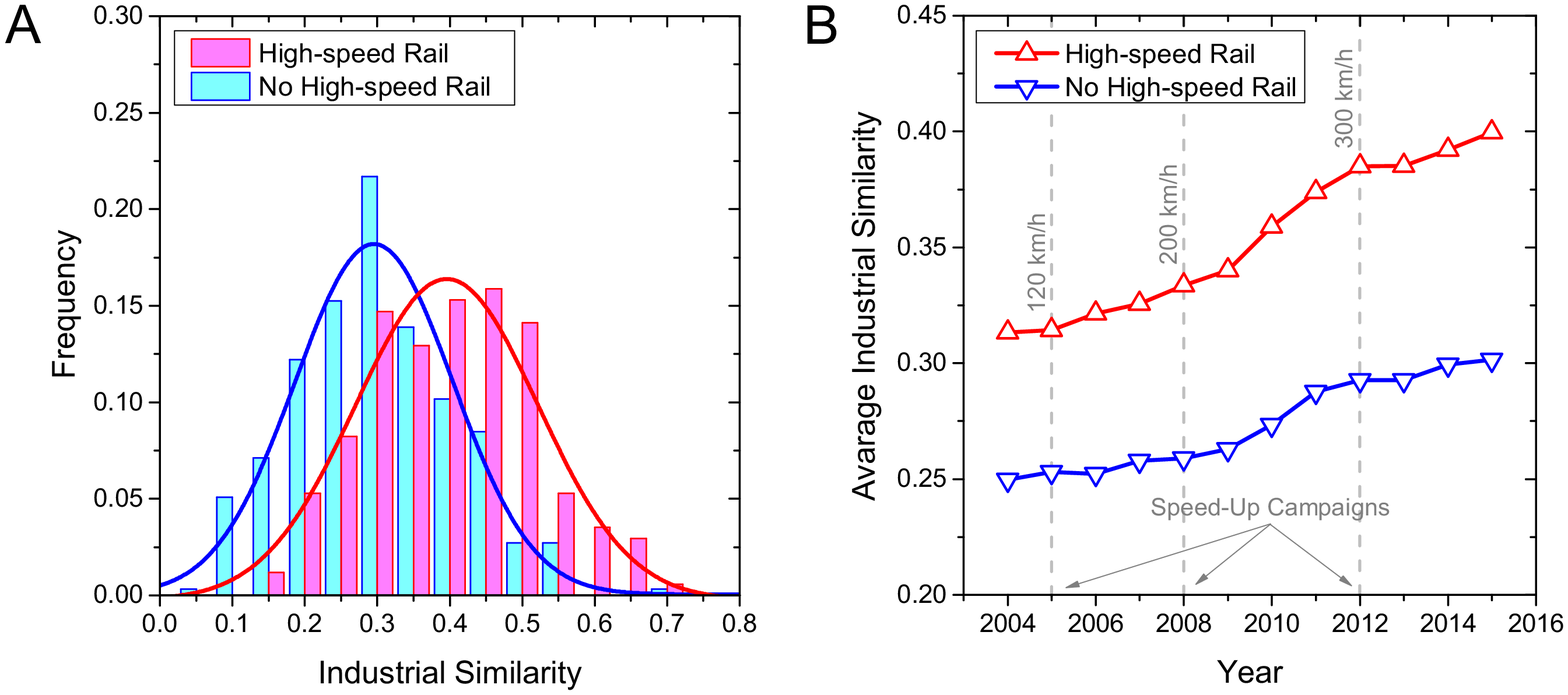}
  \caption{(A) Density distributions of industrial similarity for province pairs with (in red) or without (in blue) high-speed rail. Red and blue curves are normal fits of the bar charts. The mean of the pink distribution is significantly larger than that of the blue distribution (ANOVA p-value=$1.2\times10^{-18}$). (B) Average industry similarity between province pairs with (in red) or without (in blue) high-speed rail.}
  \label{FigS12:NeigSim}
\end{figure}

Figure~\ref{FigS12:NeigSim}A shows the effect of high-speed rail entry on industrial similarity by comparing province pairs with (in pink) and without (in blue) high-speed rail lines. We find that the average value of industrial similarity between province pairs that connected by high-speed rail is significant larger (ANOVA p-value=$1.2\times10^{-18}$). Figure~\ref{FigS12:NeigSim}B shows the timing of high-speed rail entry in China and its effect on the industrial similarity of province pairs. We can see that the average industrial similarity increases remarkably after train speed-up in 2005, 2008 and 2012 (the years after ``speed-up'' campaigns are used for illustration), suggesting the positive and significant effect of high-speed rail on inter-regional learning. The average industrial similarity for province pairs with and without high-speed rail from 2004 to 2015 supports the robustness of these findings.

To provide additional evidence supporting inter-regional learning, we do differences-in-differences (DID) analysis again but considering the productivity. First, we check the relationship between the industrial similarity and productivity for all pairs of provinces. One possible concern of our analysis is that the productivity of industries is directly affected by their industrial structure since some industries may have higher productivity than others, and the industrial structure that one province has may contribute to its productivity. In that case, the analysis of industrial similarity and productivity between pairs of provinces is not repetitively.

\begin{figure}[!t]
  \centering
  \includegraphics[width=0.9\textwidth]{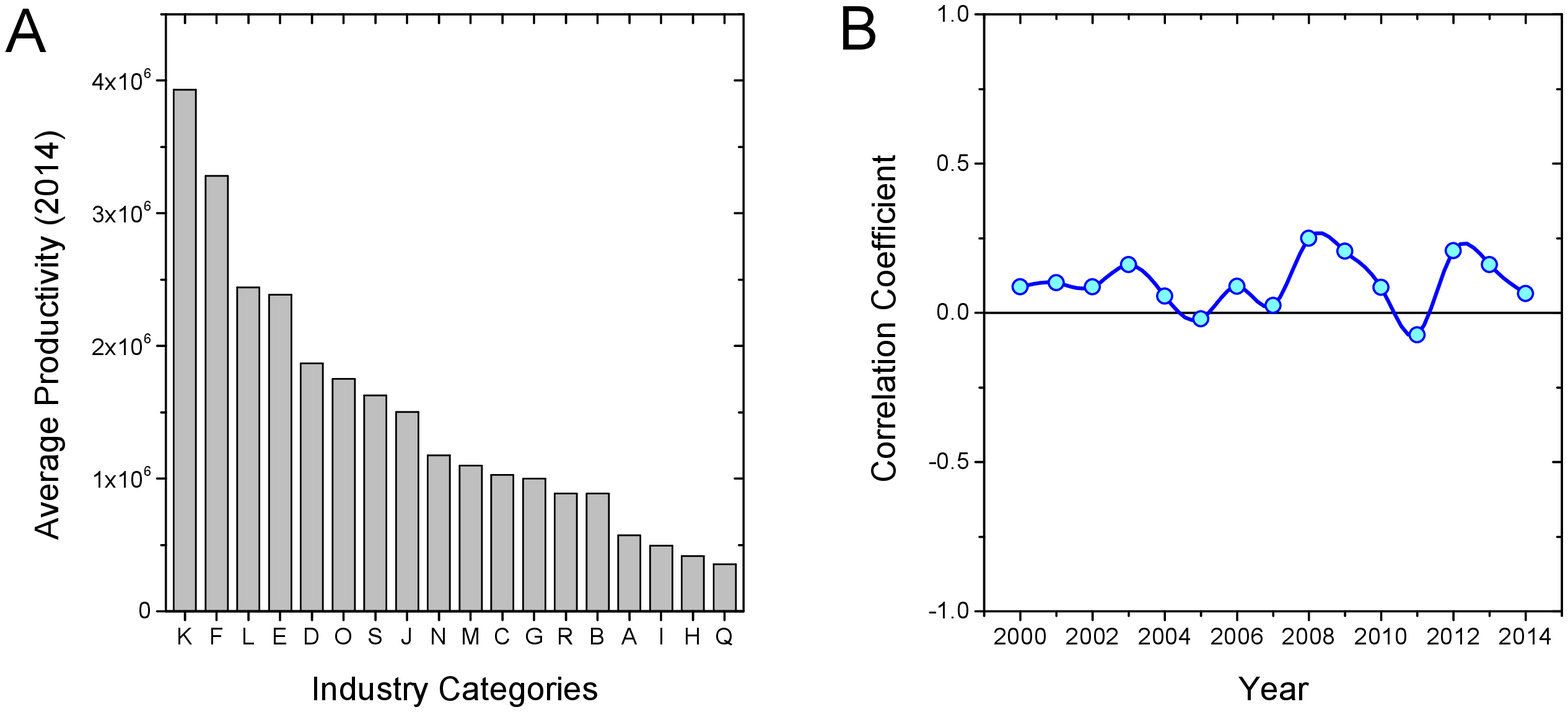}
  \caption{(A) Average productivity of industries in descending order in 2014. The keys of industry categories correspond to Figure~\ref{FigS2:CSRC}. (B) Correlation coefficient between industrial similarity and productivity for all pairs of provinces for the 2000-2014 period.}
  \label{FigS13:ProduCorr}
\end{figure}

We find that even though different industries have different productivity (see Figure~\ref{FigS13:ProduCorr}A for illustration in 2014), the correlation between industrial similarity and productivity for all pairs of provinces is relatively small (see Figure~\ref{FigS13:ProduCorr}B), meaning that industrial similarity and productivity are, to some extent, independent of each other, and considering both of these two measures at the same time is valid.

\begin{figure}[!t]
  \centering
  \includegraphics[width=0.9\textwidth]{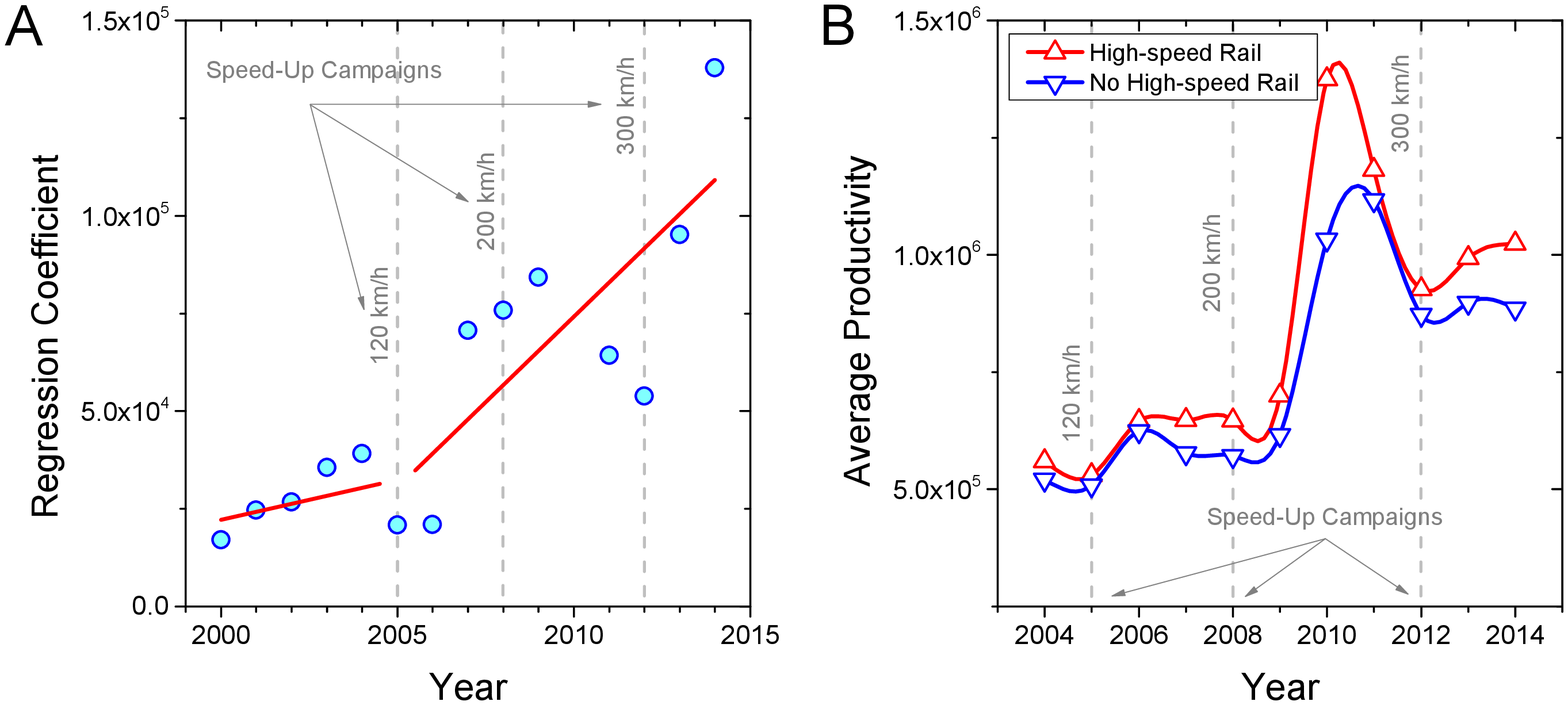}
  \caption{(A) Event study results. The $y$-axis shows the regression coefficient ($\beta_k$ in Eq.~(\ref{Eq:EventP})) as a function of the year, after regressing the average productivity of pairs of provinces that were eventually connected by high-speed rail against the entry of high-speed rail. Red lines are linear fits for 2000-2005 and 2005-2015. respectively. (B) Average productivity of province pairs with (in red) or without (in blue) high-speed rail between 2004 and 2014.}
  \label{FigS14:EventP}
\end{figure}

For doing differences-in-differences (DID) analysis considering the productivity, once again, our data satisfies the condition for DID method as the pre-trend of the dependent variable on the control and treatment groups is similar prior to year 2005 (see Figure~\ref{FigS14:EventP}A). To demonstrate this, we do the event study by running the following ordinary least-squares (OLS) linear regression model using data between 2000 and 2015 to predict the average productivity of provinces $i$ and $j$ for each year as:
\begin{equation}
  \bar p_{i,j,\alpha,t} = \beta_{0} + \sum_{k=1997}^{2015}\beta_{k}(Treat_{i,j}*1 \{t=k\}) + \varepsilon_{i,j}.
  \label{Eq:EventP}
\end{equation}
Here $Treat_{i,j}$ is a dummy variable denoting whether provinces $i$ and $j$ are affected by high-speed rail entry, and $1 \{t=k\}$ is an event time indicator, which is equal to 1 for the year that we consider the effect of high-speed rail entry. In another way, Eq.~(\ref{Eq:EventP}) regresses the average productivity of pairs of provinces and whether there is high-speed rail connecting them. Larger regression coefficient ($\beta_{k}$) corresponds to higher productivity of pairs of provinces that connected by high-speed rail than that not.

Figure~\ref{FigS14:EventP}A illustrates the results of event study regression coefficient ($\beta_{k}$) based on productivity after taking 2005 as the baseline year. Before the high-speed rail entry (1997-2005), the effect of the treatment is similar as there is no significant trend in $\beta_k$. After the high-speed rail entry (2005-2015), however, the effect of the treatment ($\beta_k$) begins to increase significantly, meaning that the treated province pairs became more productive only after the introduction of high-speed rail. Figure~\ref{FigS14:EventP}B presents the the average productivity of province pairs that connected by high-speed rail ans these not. It can be seen that, province pairs that connected by high-speed rail have significant larger productivity for the whole considered period.
\begin{table}[!t]
    \centering
    \caption{Summary statistics of variables in difference-in-difference (DID) analysis. Mean values of industrial similarity, average productivity, and differences in population (log), GDP per capita (log), urbanization and trade (log) between province pairs before and after the entry of high-speed rail are shown.}
    \footnotesize
    \begin{tabular*}{\textwidth}{@{\extracolsep{\fill}}lccccc}
    \toprule
    \multicolumn{1}{l}{\multirow{2}[4]{*}{Independent Variables}} & \multicolumn{2}{c}{Before} & \multicolumn{2}{c}{After} & \multirow{2}[4]{*}{DID} \\
\cmidrule{2-5}          & Control & Treatment & Control & Treatment &  \\
    \midrule
    Industrial Similarity & 0.2496 & 0.3133 & 0.2994 & 0.3921 & 0.0290 \\
    Productivity & $5.21\times 10^5$ & $5.60\times 10^5$ & $8.86\times 10^5$ & $10.24\times 10^5$ & $0.99\times 10^5$ \\
    $\Delta$ Population (log) & 1.1394 & 0.7314 & 1.0953 & 0.6514 & -0.0358 \\
    $\Delta$ GDP pc (log) & 0.5717 & 0.6255 & 0.4603 & 0.4327 & -0.0814 \\
    $\Delta$ Urbanization & 0.1066 & 0.2115 & 0.1098 & 0.2151 & 0.0005 \\
    $\Delta$ Trade (log) & 2.0719 & 1.5899 & 2.2047 & 1.3863 & -0.3365 \\
    \midrule
    Observations & 295   & 170   & 295   & 170   & 930 \\
    \bottomrule
    \end{tabular*}
    \label{Tab:DStat}
\end{table}

Table~\ref{Tab:DStat} shows the summary statistics of variables that are used in the differences-in-differences (DID) analysis. In our DID design, province pairs belong to the treatment group if they are connected by high-speed rail in 2014, otherwise belong to the control group. In the DID regressions, control variables include gravity considerations: the difference between population, GDP per capita, urbanization (defined as share of urban area over the entire area of a province), and trade (defined as total exports and imports of each province).

To check the robustness of our results in Table~\ref{Tab:DID} in the main text, we additionally control for the geographic distance between provinces in the DID analysis to reduce sampling bias. As shown in Table~\ref{Tab:DID2}, we find that the estimates of the effects of high-speed rail entry are also significant and robust in the presence of controlling geographic distance, supporting the causal evidence of high-speed rail entry on inter-regional learning.

\begin{table}[!t]
  \centering
  \caption{DID regressions considering the effect of high-speed rail entry on the industrial similarity and the productivity of industries with controlling for the geographic distance between provinces.}
  \footnotesize
    \begin{tabular*}{\textwidth}{@{\extracolsep{\fill}}lcccccc}
    \toprule
    \multirow{3}[6]{*}{Independent Variables} & \multicolumn{6}{c}{DID Regressions Using OLS Model} \\
\cmidrule{2-7}          & \multicolumn{3}{l}{Industrial Similarity} & \multicolumn{3}{l}{Productivity} \\
\cmidrule{2-7}          & (1)   & (2)   & (3)   & (4)   & (5)   & (6) \\
    \midrule
    \multirow{2}[1]{*}{High-speed Rail Entry} & \multicolumn{1}{l}{0.0290*} & \multicolumn{1}{l}{0.0270*} & \multicolumn{1}{l}{0.0274*} & \multicolumn{1}{l}{98713***} & \multicolumn{1}{l}{107032***} & \multicolumn{1}{l}{105193***} \\
          & \multicolumn{1}{l}{(0.0151)} & \multicolumn{1}{l}{(0.0150)} & \multicolumn{1}{l}{(0.0151)} & \multicolumn{1}{l}{(27644)} & \multicolumn{1}{l}{(27237)} & \multicolumn{1}{l}{(26044)} \\
    \multirow{2}[0]{*}{Treatment Group} & \multicolumn{1}{l}{0.0504***} & \multicolumn{1}{l}{0.0455***} & \multicolumn{1}{l}{0.0473***} & \multicolumn{1}{l}{52294***} & \multicolumn{1}{l}{40611**} & \multicolumn{1}{l}{34883*} \\
          & \multicolumn{1}{l}{(0.0112)} & \multicolumn{1}{l}{(0.0114)} & \multicolumn{1}{l}{(0.0113)} & \multicolumn{1}{l}{(16948)} & \multicolumn{1}{l}{(17528)} & \multicolumn{1}{l}{(17923)} \\
    \multirow{2}[0]{*}{After Entry} & \multicolumn{1}{l}{0.0498***} & \multicolumn{1}{l}{0.0470***} & \multicolumn{1}{l}{0.0504***} & \multicolumn{1}{l}{364939***} & \multicolumn{1}{l}{376375***} & \multicolumn{1}{l}{361673***} \\
          & \multicolumn{1}{l}{(0.0089)} & \multicolumn{1}{l}{(0.0089)} & \multicolumn{1}{l}{(0.0089)} & \multicolumn{1}{l}{(17498)} & \multicolumn{1}{l}{(17261)} & \multicolumn{1}{l}{(16491)} \\
    \multirow{2}[0]{*}{Distance (log)} & \multicolumn{1}{l}{-0.0297***} & \multicolumn{1}{l}{-0.0261***} & \multicolumn{1}{l}{-0.0278***} & \multicolumn{1}{l}{29265**} & \multicolumn{1}{l}{24052**} & \multicolumn{1}{l}{19635*} \\
          & \multicolumn{1}{l}{(0.0069)} & \multicolumn{1}{l}{(0.0069)} & \multicolumn{1}{l}{(0.0070)} & \multicolumn{1}{l}{(11579)} & \multicolumn{1}{l}{(11401)} & \multicolumn{1}{l}{(10927)} \\
    \multirow{2}[0]{*}{$\Delta$ Population (log)} &       & \multicolumn{1}{l}{-0.0182***} &       &       & \multicolumn{1}{l}{-8902} &  \\
          &       & \multicolumn{1}{l}{(0.0048)} &       &       & \multicolumn{1}{l}{(8774)} &  \\
    \multirow{2}[0]{*}{$\Delta$ GDP per capita (log)} &       & \multicolumn{1}{l}{-0.0176**} &       &       & \multicolumn{1}{l}{106183***} &  \\
          &       & \multicolumn{1}{l}{(0.0081)} &       &       & \multicolumn{1}{l}{(17249)} &  \\
    \multirow{2}[0]{*}{$\Delta$ Urbanization} &       &       & \multicolumn{1}{l}{0.0146} &       &       & \multicolumn{1}{l}{214723***} \\
          &       &       & \multicolumn{1}{l}{(0.0130)} &       &       & \multicolumn{1}{l}{(33949)} \\
    \multirow{2}[1]{*}{$\Delta$ Trade (log)} &       &       & \multicolumn{1}{l}{-0.0049**} &       &       & \multicolumn{1}{l}{19564***} \\
          &       &       & \multicolumn{1}{l}{(0.0024)} &       &       & \multicolumn{1}{l}{(4635)} \\
    \midrule
    Observations & 930   & 930   & 930   & 930   & 930   & 930 \\
    Robust $R^2$ & 0.1826 & 0.1983 & 0.1859 & 0.5013 & 0.5245 & 0.5563 \\
    RMSE  & 0.1096 & 0.1087 & 0.1095 & 2.10$\times 10^5$ & 2.00$\times 10^5$ & 2.00$\times 10^5$ \\
    \bottomrule
    \end{tabular*}
    \begin{flushleft}
    \emph{Notes}: Data are for the year 2004 (before high-speed rail entry) and 2014 (after high-speed rail entry). Significant level: $*p<0.1$, $**p<0.05$, and $***p<0.01$.
    \end{flushleft}
  \label{Tab:DID2}
\end{table}%

\end{document}